\documentclass[pra,twocolumn,aps,superscriptaddress,longbibliography]{revtex4-1}
\usepackage{amsmath}
\usepackage{amssymb}
\usepackage{graphicx}
\usepackage[usenames,dvipsnames]{color}
\usepackage{bm}
\usepackage{times}
\usepackage{hyperref}
\usepackage{float}
\usepackage{verbatim}
\usepackage{lipsum}
\usepackage{braket}
\usepackage[T1]{fontenc}
\usepackage[latin1]{inputenc}
\usepackage[table]{xcolor}
\usepackage{gensymb}
\hypersetup{
  colorlinks=true,
  citecolor=blue,
  linkcolor=blue,
  urlcolor=blue}

\definecolor{orange(colorwheel)}{rgb}{1.0, 0.5, 0.0}
\definecolor{maroon(html/css)}{rgb}{0.5, 0.0, 0.0}
\definecolor{lightgray}{gray}{0.9}

\begin{document}
%\pagewiselinenumbers
%\linenumbers
%%%%%%%%%%%%%%%%%%%%%%%%%%%%%%%%%%%%%%%%%%%%%%%%%%%%%%%%%%%%%%%%%%%%%%%%%%%%%%%
%%%%                     Title and authors                                 %%%%
%%%%%%%%%%%%%%%%%%%%%%%%%%%%%%%%%%%%%%%%%%%%%%%%%%%%%%%%%%%%%%%%%%%%%%%%%%%%%%%

\title{Quantum quench dynamics of tilted dipolar bosons in 2D optical lattices} 

\author{Hrushikesh Sable}
\affiliation{Physical Research Laboratory,
             Ahmedabad - 380009, Gujarat,
             India}
\affiliation{Indian Institute of Technology Gandhinagar,
             Palaj, Gandhinagar - 382355, Gujarat,
             India}
\author{Deepak Gaur}
\affiliation{Physical Research Laboratory,
             Ahmedabad - 380009, Gujarat,
             India}
\affiliation{Indian Institute of Technology Gandhinagar,
             Palaj, Gandhinagar - 382355, Gujarat,
             India}
\author{Soumik Bandyopadhyay}
\affiliation{Physical Research Laboratory,
             Ahmedabad - 380009, Gujarat,
             India}
\affiliation{INO-CNR BEC Center and Department of Physics,
             University of Trento, Via Sommarive 14, I-38123 Trento, Italy}
\author{Rejish Nath}
\affiliation{Department of Physics, Indian Institute of Science Education and 
	     Research, Dr. Homi Bhabha Road, Pune 411008, Maharashtra, India}
\author{D. Angom}
\affiliation{Physical Research Laboratory,
             Ahmedabad - 380009, Gujarat,
             India}
\affiliation{Department of Physics, Manipur University,
             Canchipur - 795003, Manipur, India}

\date{\today}

\begin{abstract}
 We investigate the quench dynamics of the dipolar bosons in two dimensional
optical lattice of square geometry using the time dependent Gutzwiller method.
The system exhibits different density orders like the checkerboard and the
striped pattern, depending upon the polarization angle of the dipoles.
We quench the hopping parameter across the striped density wave (SDW) to
striped supersolid (SSS) phase transition, and obtain the scaling laws for the
correlation length and topological vortex density, as function of the quench 
rate. The results are reminiscent of the Kibble-Zurek mechanism (KZM). We also 
investigate the dynamics from the striped supersolid phase to the checkerboard 
supersolid phase, obtained by quenching the dipole tilt angle $\theta$. 
This is a first order structural quantum phase transition, and we study the 
non-equilibrium dynamics from the perspective of the KZM. In particular, we 
find the number of the domains with checkerboard order follows a power law 
scaling with the quench rate. This indicates the applicability of the KZM
to this first order quantum phase transition.
\end{abstract}
\maketitle

%%%%%%%%%%%%%%%%%%%%%%%%%%%%%%%%%%%%%%%%%%%%%%%%%%%%%%%%%%%%%%%%%%%%%%%%%%%%%%%
%%%%%%%%%                      Introduction                       %%%%%%%%%%%%%
%%%%%%%%%%%%%%%%%%%%%%%%%%%%%%%%%%%%%%%%%%%%%%%%%%%%%%%%%%%%%%%%%%%%%%%%%%%%%%%

\section{Introduction}
\label{intro}
 A cornerstone of our understanding of the non-equilibrium dynamics across the 
continuous phase transitions is the Kibble-Zurek mechanism (KZM). 
The KZM provides a theoretical understanding of the spontaneous symmetry 
breaking across the continuous phase transitions, and  predicts universal 
scaling laws of the relevant physical quantities, as a function of the quench 
rate \cite{kibble_76, kibble_80, zurek_85, zurek_96, delCampo_14, dziarmaga_10}.
The central idea is the breakdown of the adiabaticity near the critical point
due to the critical slowing down. This implies that however 
slowly the system is quenched across the critical point, the dynamics ceases to
be adiabatic, leading to the topological defects.
In the context of cosmology, Kibble predicted the formation of the topological 
defects as a result of local choices of broken symmetry state in the phase
transitions \cite{kibble_76, kibble_80}. 
Later, in a novel work Zurek identified the role of the critical slowing down 
in the creation of the topological defects in condensed matter systems like 
$\text{He}^4$ \cite{zurek_85, zurek_96}. He also predicted the scaling laws for 
the correlation length and the defect density as a function of the quench rate.
The prediction has been experimentally verified with the topological defects 
in the superfluid helium \cite{bauerle_96, ruutu_96}. Since then, the KZM has 
been studied in a wide variety of systems such as liquid crystals
\cite{chuang_91, bowick_94}, superconducting loops \cite{monaco_09},
ferroelectrics \cite{lin_14}. Initially developed for the classical phase
transitions, the KZM was later applied to the quantum phase transitions (QPT)s.

 In regard to the quantum phase transitions, one of the most well-studied
systems are the ultracold atomic gases loaded in optical lattices. These
systems are considered as the macroscopic quantum simulators of the condensed
matter systems \cite{lewenstein_07,gross_17}. A prototypical example of a
quantum phase transition studied using these systems is the Mott Insulator (MI)
to superfluid (SF) phase transition \cite{fisher_89, jaksch_98, greiner_02_1, 
greiner_02_2}. The ultracold atoms in optical lattice simulate the 
Bose-Hubbard model (BHM) \cite{fisher_89, jaksch_98}, which is a minimal 
model to study the physics of interacting bosons in a lattice. 
These systems are used to understand various equilibrium quantum 
phases \cite{goral_02, zhang_15, suthar_20_1, 
suthar_20_2, bai_18, bandyopadhyay_19, pal_19,iskin_11, bai_20, malakar_20},
collective excitations \cite{suthar_15, suthar_16, suthar_17, krutitsky_10, 
krutitsky_11, saito_12} etc. In the context of the KZM, there are many
works done using the system of ultracold atoms. The early investigations in the
ultracold atoms were for harmonic trapping potentials \cite{morsch_06, 
weiler_08, lamporesi_13, donadello_16, sadler_06, damski_09, damski_10, 
anquez_16, beugnon_17}. The experimental investigations of the dynamics of the
QPT from the MI phase to the SF phase have been studied
\cite{chen_11,braun_15, cui_16}. On the theoretical front, there have been
works in investigating the evolution of the ground states during the quench
process \cite{dziarmaga_05, zurek_05, polkovnikov_05, dziarmaga_10, eckardt_05,
green_05, shimizu_misf_18, shimizu_dwss_18, zhou_20}.

 The first order phase transitions are different from the second order phase 
transitions. They are characterized by the co-existence of
multiple phases at the transition point. A natural extension of the KZM physics
is to check its applicability to the first order phase transitions. Following
this, there are few works \cite{shimizu_dwsf_18, coulamy_17} which have
studied the scaling of the topological defects in quenches across the first
order phase transitions. And recently, there are experimental works
\cite{qiu_20, liang_21} that have shown the existence of the Kibble-Zurek (KZ)
scaling across the first order phase transitions, in spinor condensates.
Motivated by this, we also investigate the quench dynamics 
across a first order phase transition, in the system of dipolar bosons in 2D 
square lattice.

 In this work, we consider the dipolar bosons in the two-dimensional square
optical lattice.  The anisotropy of the dipolar interactions between the bosons
induce different density orders like checkerboard and striped in the system,
depending on the polarization of the dipoles. To validate the predictions
of the KZM for the dipolar bosons, we first consider quenching the hopping 
amplitude across the striped density wave to striped supersolid. We investigate 
the scaling laws, predicted in the KZM, for this second order quantum phase 
transition, and obtain the critical exponents of the correlation length and the 
topological defects. Then, the polarization of the dipoles is quenched so that 
the system is driven from the striped supersolid to checkerboard supersolid 
phase. We show the existence of the impulse and adiabatic domains
in the dynamics across this first order phase transition, and calculate the
number of domains of the checkerboard order generated during the quench. This
number obeys a universal power-law scaling with the quench rate, an example
of the KZM.

 We have organized the remainder of this article as follows. In
Sec. \ref{model} we describe the model considered for our study and in the
Sec. \ref{th_methods} we discuss about the theoretical methods we have used.
We provide a description of the single-site mean-field theory used
to obtain the equilibrium ground state phases and the quenched state.
The results from the quench dynamics of the hopping parameter $J$ and the
tilt angle $\theta$ are then presented in Sec. \ref{result}.
In Sec. \ref{conclude}, we summarize the key results.

%%%%%%%%%%%%%%%%%%%%%%%%%%%%%%%%%%%%%%%%%%%%%%%%%%%%%%%%%%%%%%%%%%%%%%%%%%%%%%%
%%%%%%%   Section: Theoretical model                                %%%%%%%%%%%
%%%%%%%%%%%%%%%%%%%%%%%%%%%%%%%%%%%%%%%%%%%%%%%%%%%%%%%%%%%%%%%%%%%%%%%%%%%%%%%

\section{Theoretical Model}
\label{model}
\begin{figure}[ht]
  \includegraphics[width = 9.5cm]{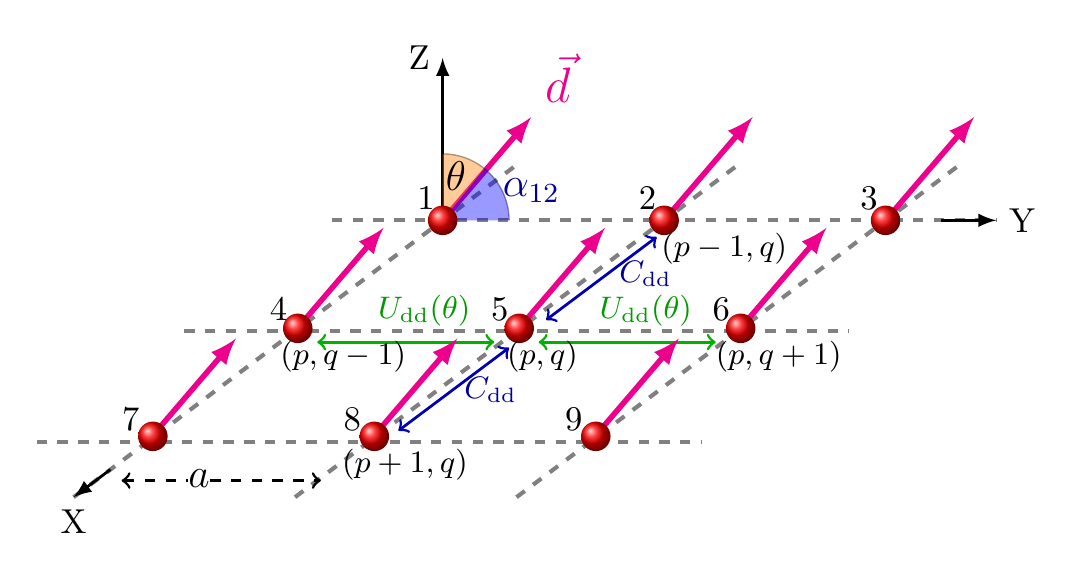}
  \caption{The schematic of the dipolar bosons in 2D optical lattice. The 
	lattice is in the $x$-$y$ plane, and the 
	polarization of the dipoles is in the $y$-$z$ plane. The 
	angle between the direction of the dipole moment $\vec{d}$ 
	and the $z$ axis is the
	tilt angle $\theta$, as is illustrated by the orange-shaded color. 
	The dipolar interaction between the nearest neighbouring dipoles along 
	the $x$ direction is represented by $C_{\rm dd}$, while the 
	interaction along the $y$ direction is given by $U_{\rm dd}(\theta)$.}
  \label{schmatic}
\end{figure}

We consider a system of dipolar bosons in 2D optical lattice of square
geometry with the lattice constant $a$.
The lattice plane constitutes the $x$-$y$ plane of the chosen co-ordinate system
of the model, as shown schematically in the Fig.~\ref{schmatic}. At zero 
temperature, the physics of the system is well described by the lowest band BHM 
with the dipolar interaction and the Hamiltonian of the system  
is \cite{goral_02, yi_07, danshita_09, bandyopadhyay_19, wu_20}  
\begin{eqnarray}
  \hat{H} &= &-\sum_{\langle i,j \rangle}J\left( \hat{b}_{i}^{\dagger}
               \hat{b}_{j} + {\rm H.c.} \right)
              + \sum_{i}\hat{n}_{i} \left [\frac{U}{2}(\hat{n}_{i}
              -1)- \mu \right]  \nonumber \\ 
          && +  \frac{C_{\rm dd}}{2}\sum_{ij}\hat{n}_i\hat{n}_j
	     \frac{(1-3\text{cos}^2\alpha_{ij})}{|\mathbf{r}_i 
             - \mathbf{r}_j|^3},
  \label{dipbhm}
\end{eqnarray}
where $i \equiv (p,q)$ represent the lattice indexes, and 
$j \equiv (p',q')$ are it's nearest neighbouring lattice site indexes,
$\hat{b}_{i}^{\dagger}$ ($\hat{b}_{i}$) are the creation (annihilation)
operators, $\hat{n}_{i}$ is the bosonic occupation number operator and the
summation indexes within $\langle\cdots\rangle$ denote the sum over the
nearest neighbours. Further, $J$ is the hopping strength, $U>0$ is the
on-site inter-atomic interaction strength, and $\mu$ is the chemical
potential. The coupling constant $C_{\rm dd} \propto d^{2}/a^{3}$ represents 
the strength of the dipolar interaction. 
Here, $d$ represents the magnitude of either the permanent 
magnetic dipole moment, which atoms like Cr, Er and Dy possess, 
or electric dipole moment in the case of polar molecules.
The angle $\alpha_{ij}$ 
is the angle between the polarization axis and the separation vector 
$\mathbf{r}_{i} - \mathbf{r}_{j}$. The long range and the anisotropic nature 
of the dipolar interaction induce various symmetry broken phases.
In our study, for simplicity, we consider the model where the dipolar 
interaction is restricted to the NN sites. This is an optimal model which
encapsulates the physics of the long range dipolar interactions of ultracold
atoms in a lattice.
%%%
Then the Hamiltonian is 
\begin{eqnarray}
  \hat{H} &= &-\sum_{\langle i,j \rangle}J\left( \hat{b}_{i}^{\dagger}
              \hat{b}_{j} + {\rm H.c.} \right)
              + \sum_{i}\hat{n}_{i} \left [\frac{U}{2}(\hat{n}_{i}
              -1)- \mu \right]  \nonumber \\ 
          &&  +  \frac{C_{\rm dd}}{2}\sum_{\langle ij \rangle}
              \hat{n}_i\hat{n}_j (1 - 3\text{cos}^2\alpha_{ij}).
  \label{nndipbhm}
\end{eqnarray} 
The dipoles are assumed to be polarized in the $y$-$z$ plane, as demonstrated
in Fig.~\ref{schmatic}. The angle between the $z$-axis and the polarization 
axis is denoted by $\theta$, and is referred to as the tilt angle. The tilt 
angle can be experimentally controlled by applying an external electric and 
magnetic field for the electric and magnetic 
dipoles, respectively. In this work, we consider the atoms with magnetic 
dipole moment, as these are primarily used in experiments on 
dipolar Bose-Einstein condensates (BECs) \cite{baier_16}. Since the 
polarization is in the $y$-$z$ plane, it should be noted that 
$\theta = \pi/2 - \alpha_{ij}$. The isotropic limit of the model is
obtained when $\theta = 0$, and this is commonly referred to as the 
extended BHM (eBHM) \cite{iskin_11, suthar_20_1}. For the configuration 
considered, the dipole-dipole interaction along the $x$-axis is always 
repulsive and independent of $\theta$. Along the $y$-axis the 
interaction is $\theta$
dependent and it has form 
$U_{\rm dd}(\theta) = C_{\rm dd}(1 - 3\sin^{2}\theta)$. 
The interaction along the $y$-axis is repulsive when $\theta < \theta_{\rm M}$, 
and attractive for $\theta > \theta_{\rm M}$, where 
$\theta_{\rm M} = \sin^{-1}\left(1/\sqrt{3}\right) \approx 35.3\degree$, 
is referred to as the magic angle. At the magic angle, the dipole 
interaction along the $y$-axis is zero. 
In the present work, we first focus on
the case $\theta > \theta_{\rm M}$, and study the interplay of the attractive
$U_{\rm dd}(\theta)$ with the repulsive $U$ and the dipolar interaction
along $x$-axis. Then, we tune the tilt angle $\theta$ across the magic angle
$\theta_M$ and study the structural phase transition between the striped
supersolid phase and the checkerboard supersolid phase.

%%%%%%%%%%%%%%%%%%%%%%%%%%%%%%%%%%%%%%%%%%%%%%%%%%%%%%%%%%%%%%%%%%%%%%%%%%%%%%%
%%%%%%%   Section:  Theoretical methods                             %%%%%%%%%%%
%%%%%%%%%%%%%%%%%%%%%%%%%%%%%%%%%%%%%%%%%%%%%%%%%%%%%%%%%%%%%%%%%%%%%%%%%%%%%%%

\section{Theoretical Methods}
\label{th_methods}
\subsection{Single site Gutzwiller mean-field (SGMF) theory}
 We use the SGMF method to obtain the ground state of the model. In this 
method, the annihilation (creation) operators at lattice site
$(p,q)$ in Eq.(\ref{nndipbhm}) are decomposed as \cite{rokhsar_91, 
sheshadri_93, bai_18, pal_19, bandyopadhyay_19, suthar_20_1, bai_20, 
suthar_20_2, malakar_20} 
\begin{subequations}
\begin{eqnarray}
  \hat{b}_{p, q}           &=& \phi_{p,q} + \delta \hat{b}_{p, q},       \\
  \hat{b}^{\dagger}_{p, q} &=& \phi^{*}_{p, q} + \delta \hat{b}^{\dagger}_{p, q}
 \label{decompose} 
\end{eqnarray}
\end{subequations}
where, $\phi_{p,q} = \langle\hat{b}_{p,q}\rangle$, and $\phi^{*}_{p, q} = 
\langle\hat{b}^{\dagger}_{p,q}\rangle$ are the mean field and its complex
conjugate, respectively. Here the expectation values are taken with respect
to the ground state of the system. A similar application is done to the 
number operator $\hat{n}_{p,q}$ and then, we can write the single-site mean
field Hamiltonian as
\begin{equation}
   \hat{h}_{p,q} = \hat{h}^{\rm BHM}_{p,q} + \hat{h}^x_{p,q} 
                   + \hat{h}^y_{p,q},
\end{equation}
where, the single-site Bose-Hubbard model Hamiltonian is 
\begin{eqnarray}
  \hat{h}^{\rm BHM}_{p,q} &=& -J \bigg [\Big(\phi_{p + 1, q}^{*}
		      \hat{b}_{p, q} -  
                      \phi_{p+1,q}^{*}\phi_{p, q} + \nonumber \\ 
		   &&  \phi_{p, q+1}^{*}\hat{b}_{p, q}  
		    - \phi_{p, q+1}^{*}\phi_{p, q}\Big)
                    + {\rm H.c.}\bigg ] \nonumber \\
                 && +\frac{U}{2}\hat{n}_{p, q}(\hat{n}_{p, q}-1) - \mu_{p, q}
                      \hat{n}_{p, q},
  \label{mf_bhm}
\end{eqnarray}
the single-site mean field Hamiltonian arising from the long range interaction
along $x$-axis is 
\begin{equation}
  \hat{h}^x_{p,q} = \frac{C_{\rm dd}}{2}\sum_{p'}
		    \Big(\langle\hat{n}_{p',q}\rangle 
                    \hat{n}_{p,q} - \langle\hat{n}_{p',q}\rangle
                    \langle \hat{n}_{p,q}\rangle\Big),
  \label{mf_xdd}
\end{equation}
with $p'=p-1$ and $p+1$. Similarly, for the long range interaction along the 
$y$-axis, the single-site mean field Hamiltonian is 
\begin{equation}
  \hat{h}^y_{p,q} = \frac{U_{\rm dd}(\theta)}{2}\sum_{q'}
	            \Big(\langle\hat{n}_{p,q'}\rangle
                    \hat{n}_{p,q}- \langle\hat{n}_{p,q'}\rangle
                    \langle \hat{n}_{p,q}\rangle\Big),
  \label{mf_ydd}
\end{equation}
with $q'=q-1$ and $q+1$. The total mean field Hamiltonian of the system is 
then,
\begin{equation}
  \hat{H}^{\rm MF}= \sum_{p, q} \hat{h}_{p,q}.
  \label{h_mf}
\end{equation}
Using the Gutzwiller ansatz, the ground state of the system is 
\begin{equation}
 \ket{\Psi_{\rm GW}} = \prod_{p, q}\ket{\psi}_{p, q}
                     = \prod_{p, q} \sum_{n = 0}^{N_{\rm b}}c^{(p,q)}_n
                       \ket{n}_{p, q},
 \label{gutzwiller_wv_fn}
\end{equation}
where, $N_b$ is the maximum allowed occupation number basis (Fock space 
basis), and $c^{(p,q)}_n$ are the coefficients of the occupation number 
state $\ket{n}$ at the lattice site $(p,q)$. The SF mean-field order 
parameter $\phi_{p, q}$ is calculated as
\begin{equation}
  \phi_{p, q} = \langle\Psi_{\rm GW}|\hat{b}_{p, q}|\Psi_{\rm GW}\rangle 
              = \sum_{n = 1}^{N_{\rm b}}\sqrt{n} 
                {c^{(p,q)}_{n-1}}^{*}c^{(p,q)}_{n}.
\label{gw_phi}             
\end{equation}
The occupancy at a lattice site $(p, q)$ is the expectation of the number 
operator
\begin{equation}
   \langle \hat{n}_{p,q}\rangle = 
    \sum_{n = 0}^{N_{\rm b}} | c_n^{(p,q})|^2 n_{p,q}.
   \label{number} 
\end{equation}
Since the mean field Hamiltonian depends on $\phi$ and $\langle \hat{n} 
\rangle$, which are the expectation values of single-site operators, 
we initialize their values by choosing a specific local wavefunction. 
In general, we choose 
the single site wavefunction such that $c_n^{(p,q)}$ have the same magnitude,
and normalize it. We, then, solve for the ground state of each site by 
diagonalizing the corresponding single site Hamiltonian. And, recalculate 
the SF order parameter and the expectation of the number operator. This is 
done for all of the lattice sites and one such sequence of calculation 
constitutes one iteration. The iterations are repeated till we achieve
convergence.

%%%%%%%%%%%%%%%%%%%%%%%%%%%%%%%%%%%%%%%%%%%%%%%%%%%%%%%%%%%%%%%%%%%%%%%%%%%%%%%
%%%%%%%   Subsection:Characterization of quantum phases             %%%%%%%%%%%
%%%%%%%%%%%%%%%%%%%%%%%%%%%%%%%%%%%%%%%%%%%%%%%%%%%%%%%%%%%%%%%%%%%%%%%%%%%%%%%

\subsection{Characterization of quantum phases}
 The quantum phases of the dipolar BHM are identified by a unique combination 
of different order parameters. The ground state of the standard BHM 
exhibits two quantum phases: Mott insulator (MI) and the superfluid (SF) 
phase. The MI phase is an insulating phase, and has fixed, integer 
commensurate lattice site occupancies. The SF phase is a compressible phase, 
and displays the characteristic off-diagonal long range order (ODLRO). The 
SF order parameter $\phi$ is zero in the MI phase, and is finite in the SF 
phase. Thus,  $\phi$ is an order parameter which distinguish the two quantum
phases of the standard BHM. For ultracold dipolar atoms in an optical lattice,
the long range interactions introduce additional quantum phases. These are
structured phases like the density-wave (DW) and the supersolid (SS) phases. 
These phases have diagonal long-range crystalline order. The SS phase, in 
addition, has ODLRO. Hence, it has non-zero $\phi$ and periodic modulation 
of $\langle \hat{n}_{p,q} \rangle$. The periodic modulation of the density is 
captured by the structure factor $S(\mathbf{k})$ \cite{bandyopadhyay_19}, 
where $\mathbf{k} = k_x \hat{i} + k_y \hat{j}$ is the two-dimensional 
reciprocal lattice vector,
\begin{equation}
   S(\mathbf{k}) = \frac{1}{N^{2}}\sum_{i,j} e^{i\mathbf{k}.(\mathbf{r}_i 
                   - \mathbf{r}_j)} \langle \hat{n}_i \hat{n}_j \rangle,
   \label{strucfact}
\end{equation}
and $N$ is the total number of bosons in the lattice. As mentioned earlier,
we first consider $\theta >\theta_M$ domain and study the hopping quenching. 
Then the associated anisotropy of the long range interaction induces the 
periodic density modulations along the $x$-axis, since the interaction along 
the $y$-axis is attractive. Hence, for $(k_{x}, k_{y}) = (\pi, 0)$, 
$S(\mathbf{k})$ is  non-zero in the DW and the SS phases. This implies that 
the translational symmetry is broken along the $x$ direction. Henceforth, we 
shall be referring to these quantum phases as striped density wave SDW 
($n_A, n_B$) and striped supersolid (SSS) phases in our study. 
Here, $n_A$ and $n_B$ are the occupancies of the 
two consecutive lattice sites along the $x$ direction. For $\theta < \theta_M$, 
the dipolar interaction induces periodic density modulation along both the
$x$ and $y$ directions. The quantum phases then have checkerboard order, and 
are referred as checkerboard density wave (CBDW) and checkerboard 
supersolid (CBSS) phases. We use $S(\pi,\pi)$ to identify these ordered phases.
This implies that these checkerboard phases break the translational symmetry
along both the directions.
Table~\ref{tab_phase} summarizes
the classification of all the quantum phases discussed in the present work.
\begin{table}[ht]
  \begin{tabular}{l | c | c | c | c }
   \hline \hline
     Quantum phases &${n}_{p,q}$ & $\phi_{p,q}$ & $S(\pi,0)$ & $S(\pi,\pi)$\\
   \hline
     Mott insulator (MI)  & Integer & $0$      & $0$  &$0$\\
     Striped Density wave (SDW)    & Integer & $0$      & $\neq 0$ & $0$ \\
     Striped Supersolid (SSS)      & Real    & $\neq 0$ & $\neq 0$ & $0$  \\
  Checkerboard Density wave (CBDW) & Integer  & $ 0$   & $0$   & $\neq 0$  \\
  Checkerboard Supersolid (SSS)    & Real    & $\neq 0$ & $0$ & $\neq 0$  \\
     Superfluid (SF)      & Real    & $\neq 0$ & $0$ &$0$  \\
   \hline
  \end{tabular}
  \caption{Classification of quantum phases based on different order
           parameters.}
  \label{tab_phase}
\end{table}

%%%%%%%%%%%%%%%%%%%%%%%%%%%%%%%%%%%%%%%%%%%%%%%%%%%%%%%%%%%%%%%%%%%%%%%%%%%%%%%
%%%%%%%   Subsection:Quench dynamics and KZ scaling relations       %%%%%%%%%%%
%%%%%%%%%%%%%%%%%%%%%%%%%%%%%%%%%%%%%%%%%%%%%%%%%%%%%%%%%%%%%%%%%%%%%%%%%%%%%%%

\subsection{Quench dynamics and KZ scaling relations.}

In this work we study the dynamics of the system when a chosen parameter
$\chi$  undergoes quantum quench from an initial value 
$\chi_i$ to a final value $\chi_f$, here $\chi$ is either $J$ or $\theta$.
During the quench the parameter is a function of time $\chi(t)$
and modifies the Hamiltonian in Eq.(\ref{h_mf}) to a time dependent one. 
The other parameters $C_{\rm dd}$, $U$ and $\mu$ remain fixed or time 
independent. Then, the time-dependent Schrodinger equation 
\begin{equation}
   i\hbar\partial_t \Ket{\psi}_{p,q} = \hat{h}_{p,q} \Ket{\psi}_{p,q},
   \label{schdinger_eqn}
\end{equation}
describes the temporal evolution of the single site wavefunction.
This results to a set of coupled, linear partial differential equations 
of the coefficients $c_n^{(p,q)} (t)$. To solve these equations we employ the 
fourth-order Runge-Kutta method and once we have $c_n^{(p,q)} (t)$ the 
wavefunction of the system at a particular instant of time $t$ is defined. 
To initiate the quantum quench, we obtain the equilibrium wavefunction with 
the $\chi = \chi_i$. 
In the next step of the state preparation for the quantum 
quench dynamics, we introduce fluctuations to the coefficients 
$c_n^{(p,q)}$ of this state. For the $J$ quenching, we introduce phase and 
density fluctuations. The phase fluctuations are introduced to the non-zero 
coefficients of the wavefunction. For this we generate uni-variate random 
phases in the domain $[0, 2\pi]$. We, then, introduce density fluctuations 
by adding noise to the amplitudes of the coefficients $c_n^{(p,q)}$. This is 
done by generating uni-variate random numbers in the domain $[0, \Delta]$, with 
$\Delta \approx 10^{-4}$. In the case of the $\theta$ quenching, the density
fluctuations alone is sufficient to drive the SSS-CBSS transition as it 
involves a change in the density order. The choice of the strength 
$\Delta$ is based on a detailed analysis of the quantum and thermal 
fluctuations using Bogoliubov-de Gennes analysis of the collective modes, see 
Appendix \ref{bdg_appendix}. These fluctuations simulate the effects of the 
quantum fluctuations in the system, and are essential to drive the quantum 
phase transitions. To obtain reliable statistics, we consider an ensemble 
consisting of 80 such randomized initial states. 
And, each of these states are evolved in time by quenching the 
appropriate parameter. 
Then, a measure of a physical observable 
is the ensemble average over all the 80 samples. Furthermore, for each 
member of the ensemble we consider the average value of the observable 
across the entire system. For example, to examine the quench dynamics of the 
SF order parameter, we consider the system average 
\begin{align}
  |\Phi| = \frac{\sum_{p,q}|\phi_{p,q}|}{N_s},
\end{align}
as a representative value and here, $N_s$ is the total number of 
lattice sites in the system.

There are several important temporal markers in a quantum quench dynamics. 
The Kibble-Zurek mechanism categorizes the entire quench process into three
temporal regimes \cite{damski_05,damski_06}. Assuming that the quench is 
linear and is initiated at time $t = -\tau_Q$,
the time duration $[-\tau_Q, -\hat{t}]$ marks the adiabatic regime. As the 
name suggests, in this temporal regime, the relaxation time $\tau$ of the 
system is short and the system adjusts to the temporal change in the 
parameter. 
In other words, $\tau$ is smaller than the time 
scale on which the quench is performed. As the system approaches the quantum
critical point at $t=0$, $\tau$ diverges, and dynamics is frozen. That is,
the state of the system cannot follow the change in the parameter. 
This scenario persists till the time $\hat{t}$, and the time duration  
$[-\hat{t}, \hat{t}]$ marks the impulse period of the quench. The time 
instant $\hat{t}$ is generally referred to as the transition time, and the 
rate of the quench is $1/\tau_Q$. For $t > \hat{t}$, the system is again in 
the adiabatic regime.

 The Kibble-Zurek theory \cite{kibble_76, kibble_80,zurek_85, zurek_96, 
delCampo_14} predicts the rate of the topological defects formation during 
the course of the non-equilibrium dynamics. These defects are generated due to 
the different local choices of the order parameter associated with the 
symmetry broken phase. The symmetry breaking occurs spontaneously and 
the system undergoes a quantum phase transition  as it crosses the 
critical point at $t=0$. For an extended system, the choice of the broken 
symmetry must propagate across the entire system for it to be uniform. 
But, depending on the rate governed by $\tau_Q$, this is forbidden, and 
finite sized domains are formed in the system. In other words, the formation 
of these domains is a direct consequence of causality \cite{zurek_96}. 
In the case of the $J$ quenching, which results in 
spontaneous breaking of global $U(1)$ symmetry, the defects manifest as 
vortices, and we compute their 
density as follows \cite{shimizu_misf_18, shimizu_dwss_18, shimizu_dwsf_18}:
\begin{equation}
  N_v = \sum_{p,q} |\Omega_{p,q}|, 
  \label{vort_den}
\end{equation}
with
\begin{eqnarray}
  \!\!\!\!\!\!\!\!
  \Omega_{p,q} &=& \frac{1}{4}\big [\sin(\theta_{p+1,q} - \theta_{p,q})
                   + \sin(\theta_{p+1,q+1} - \theta_{p+1,q})  
                                         \nonumber \\
               &&  -\sin(\theta_{p+1,q+1} - \theta_{p,q+1}) - 
                   \sin(\theta_{p,q+1} - \theta_{p,q})\big].
  \label{vort_def}
\end{eqnarray}
Here, $\theta_{p,q}$ is the phase of the SF order parameter $\phi_{p,q}$.
For large $\tau_Q$, the quench rate is small, and the number of the 
vortices $N_v$ is small. Thus, $N_v$ satisfies the scaling law 
\begin{equation}
   N_v \propto \tau_{Q}^{-d},
   \label{nv-d}
\end{equation}
where, $d$ is the critical exponent. We also compute the correlation length
$\xi$ \cite{shimizu_misf_18, shimizu_dwss_18, shimizu_dwsf_18},
which captures the correlation between the lattice sites of the system.
It is defined as 
\begin{equation}
  \langle b_{i}^{\dagger} b_{j}\rangle \propto 
   \exp\bigg(\frac{-|i-j|}{\xi}\bigg).
\end{equation}
For a given pair $\{i,j\}$ of lattice sites, we invert the above relation
to calculate $\xi$ for the pair. We then average this quantity over all 
possible pairs in the lattice, and this also covers all possible spatial
separations. The scaling law of $\xi$ is
\begin{equation}
   \xi \propto \tau_{Q}^b ,\nonumber 
\end{equation}
where, $b$ is the critical exponent. The critical exponents $b$ and $d$ 
are related to the other critical exponents $\nu$ and $z$ as 
\begin{subequations}
\begin{eqnarray}
   b &=& \frac{\nu}{1 + \nu z}  \\
   d &=& \frac{2\nu}{1 + \nu z}. 
\end{eqnarray}
\label{crit_exp}
\end{subequations}
Here, $\nu$ is the critical exponent of the equilibrium correlation length 
and $z$ is the dynamical critical exponent. Thus, the KZ mechanism predicts 
the scaling relation $d = 2b$. In the present work, one of our aim is to 
validate this scaling relation for the quantum phase transition from the 
SDW phase to SSS phase. 
For the $\theta$ quenching across the SSS-CBSS 
transition, domains of the checkerboard order emerge during the dynamics. 
Analogous to the defect density, we quantify the number of domains $N_D$, 
and obtain a power-law scaling of it with the quench rate.

%%%%%%%%%%%%%%%%%%%%%%%%%%%%%%%%%%%%%%%%%%%%%%%%%%%%%%%%%%%%%%%%%%%%%%%%%%%%%
%%%%%% Section:  Results                                              %%%%%%%
%%%%%%%%%%%%%%%%%%%%%%%%%%%%%%%%%%%%%%%%%%%%%%%%%%%%%%%%%%%%%%%%%%%%%%%%%%%%%

\section{Results}
\label{result}
When the tilt angle $\theta$ is greater than $\theta_{\rm M}$, the magic 
angle, the density wave (DW) and supersolid (SS) phases have striped order 
along the $y$-axis. This configuration is energetically favourable as the
NN dipolar interaction along the $y$ direction is attractive for 
$\theta > \theta_{\rm M}$ and the dipoles are aligned head-to-tail
formation. We first consider $\theta = 40\degree$ and study the quantum quench 
from striped density wave (SDW(1,0)) to striped supersolid (SSS). 
This $\theta$ value is chosen so that the system is deep in the striped
domain, and avoids the instabilities associated with large $\theta$. We 
present the equilibrium phase diagram for the dipolar BHM at 
$\theta = 40\degree$, in the $J/U$-$\mu/U$ plane obtained using the SGMF 
method. After verifying the KZ scaling relations for the SDW(1,0) to 
SSS transition, we consider the $\theta$ quench across the
SSS-CBSS phase transition. We verify the presence of the impulse and adiabatic
domains in the dynamics, and quantify the domains of the emerging 
checkerboard order. We consider the system size as $100 \times 100$ in our 
computations. We scale the Hamiltonian with $U$ and time is defined in the
units of $\hbar/U$.

%%%%%%%%%%%%%%%%%%%%%%%%%%%%%%%%%%%%%%%%%%%%%%%%%%%%%%%%%%%%%%%%%%%%%%%%%%%%%
%%%%%% Subsection: Equilibrium phase diagram                          %%%%%%%
%%%%%%%%%%%%%%%%%%%%%%%%%%%%%%%%%%%%%%%%%%%%%%%%%%%%%%%%%%%%%%%%%%%%%%%%%%%%%
\subsection{Equilibrium phase diagram}
\begin{figure}[ht]
  \includegraphics[scale=0.6]{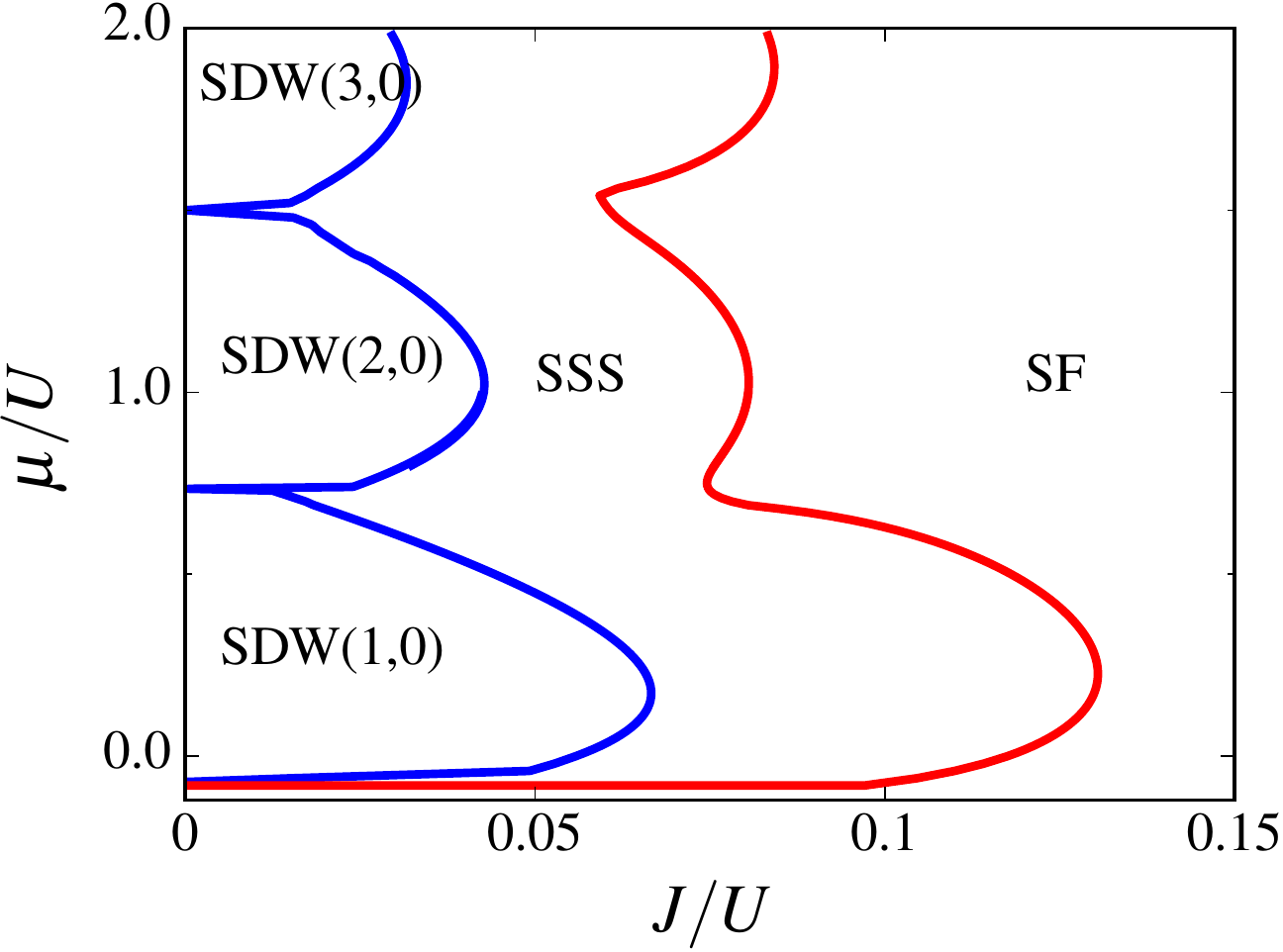}
	\caption{Phase diagram of dipolar BHM for $\theta = 40\degree$ in the 
	   $J/U - \mu/U$
	   plane. The blue line denotes the SDW - SSS phase boundary, 
	   while the red line denotes the SSS-SF phase boundary.} 
\label{ph_ss_thta40}
\end{figure}
\begin{figure}[ht]
  \includegraphics[width = 8.5cm]{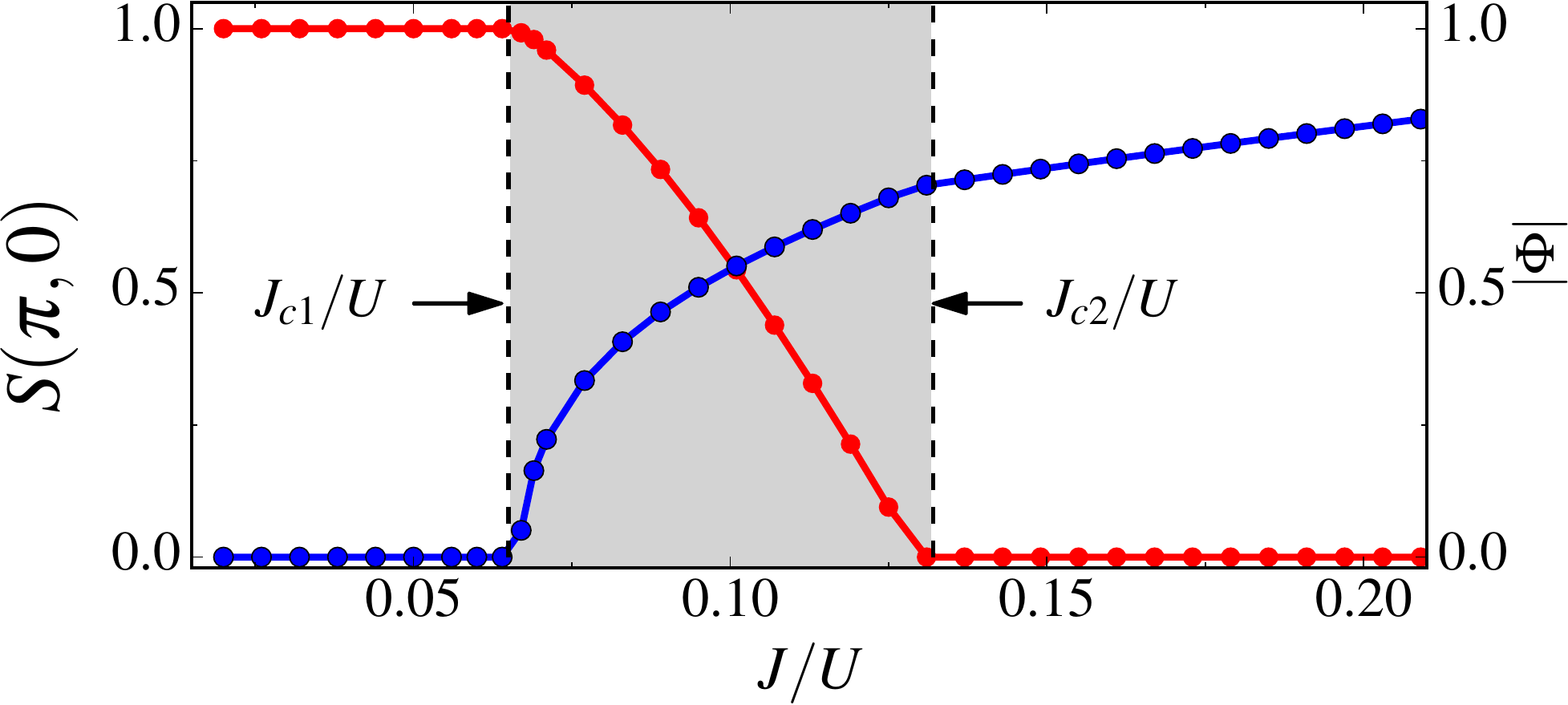}
  \caption{$S(\pi,0)$ (red) and $|\Phi|$ (blue) as a function of 
	   $J/U$, for $\mu/U = 0.17$. The black arrows mark the two transition
	   points $J_{c1}/U$ and $J_{c2}/U$, for the SDW to SSS and SSS to SF
	   phase transition, respectively.} 
\label{eq_sfpz}
\end{figure}

The ground state phase diagram of the Hamiltonian with anisotropic NN 
interaction, given by Eq.(\ref{nndipbhm}), is shown in 
Fig.~\ref{ph_ss_thta40} for $\theta = 40\degree$. At low values of $J/U$
or in the strongly interacting domain, the ground state of the system is 
in the insulating SDW  \cite{zhang_15, bandyopadhyay_19} phase. 
The SDW phase, like the MI phase, is incompressible and has $\phi = 0$. The
latter implies the absence of ODLRO in this phase. For large values of $J/U$,
the system is in the SF phase and possess ODLRO. The SF phase has
translational invariance, and hence, it has $S(\pi,0) = 0$. 
In the Fig.~\ref{eq_sfpz}, the variation of the $S(\pi,0)$ and $|\Phi|$ is
shown as a function of $J/U$ for  $\mu/U = 0.17$. When the system is in 
the SDW(1,0) phase, the value of $S(\pi,0)$ is unity and $|\Phi| = 0$. The 
domains of the two quantum phases SDW and SF are separated by the SSS phase.
In the SSS phase, both the order parameters $S(\pi,0)$ and $|\Phi|$ are 
non-zero. The structure of the occupancies in SSS phase are similar to the 
SDW phase, but, the occupancies are real. The SSS phase has both the ODLRO 
and diagonal long range order \cite{leggett_70, ohgoe_12a,zhang_15, 
bandyopadhyay_19}. Thus, the SDW and the SF phases are separated by two 
second-order phase transitions, SDW to SSS, and SSS to SF.  

In the subsequent subsection, we study the dynamics of the second-order 
quantum phase transition from the SDW(1,0) phase to SSS phase, at fixed 
$\mu/U$ and $C_{\rm dd}/U$.

%%%%%%%%%%%%%%%%%%%%%%%%%%%%%%%%%%%%%%%%%%%%%%%%%%%%%%%%%%%%%%%%%%%%%%%%%%%%%
%%%%%% Subsection: Equilibrium phase diagram                          %%%%%%%
%%%%%%%%%%%%%%%%%%%%%%%%%%%%%%%%%%%%%%%%%%%%%%%%%%%%%%%%%%%%%%%%%%%%%%%%%%%%%

\subsection{Transition from SDW to SSS} 
\label{J quenching}
\begin{figure}[ht]
  \includegraphics[width = 7.5cm]{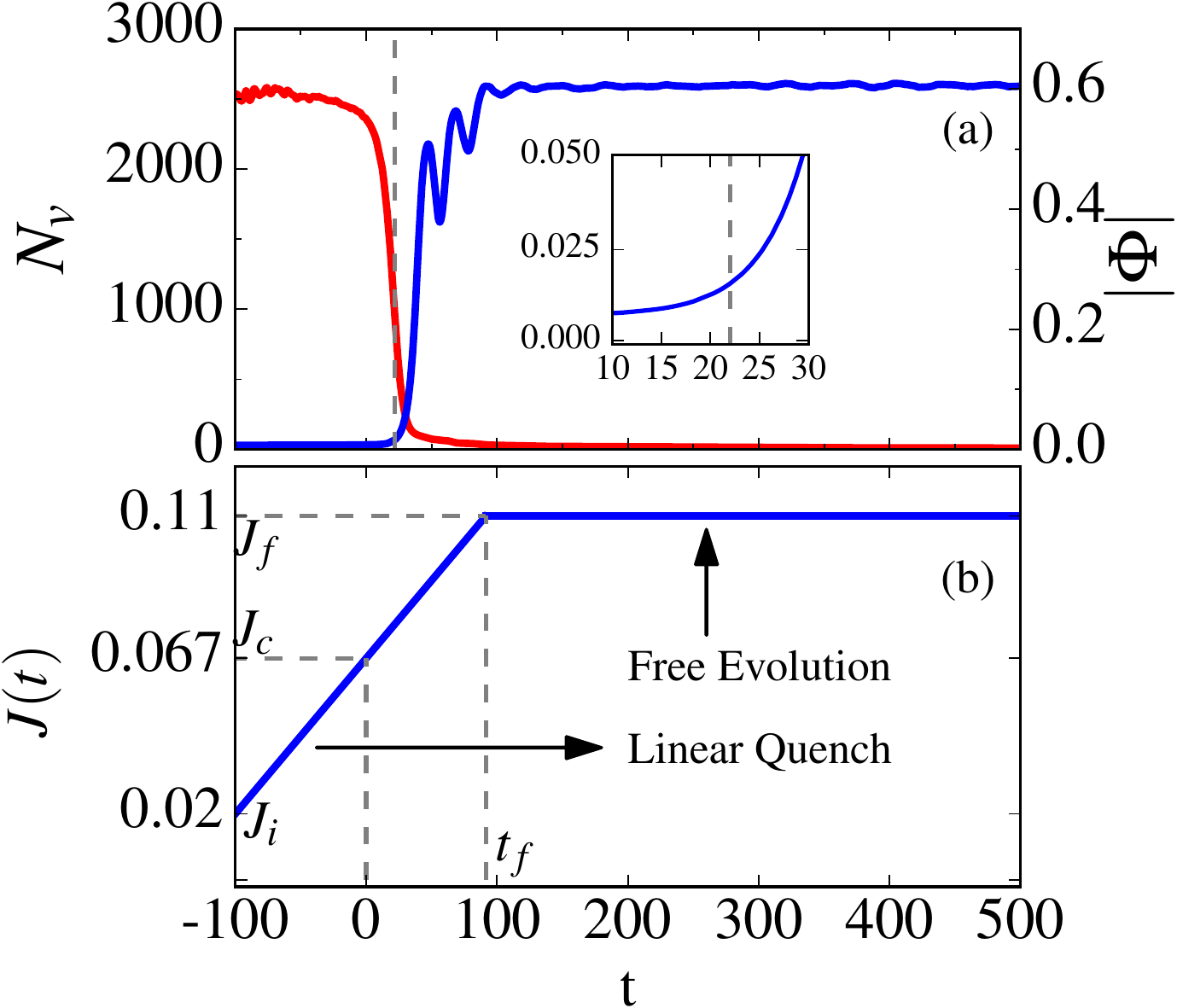}
	\caption{The time evolution of  $|\Phi|$ (blue) and 
	 $N_v$ (red) 
	 for $\tau_Q = 100$ is shown in (a). The inset in (a) depicts the 
	 behaviour of $|\Phi|$ near $\hat{t}$.
	 The critical point of the phase transition is passed at $t=0$. The
	 dashed line in (a) indicates the time $\hat{t}$, at which the SF order 
	 parameter rises steeply, as illustrated in the inset. 
	 As the system enters the SSS phase, there is an annihilation of the 
	 vortices and hence, $N_v$ decreases to zero. In (b), we schematically
	 illustrate the quench protocol used in our simulations.
	 The hopping amplitude is ramped up from value $J_i$ to $J_c$ in 
	 time $\tau_Q$ (here $\tau_Q = 100)$. After the quench is terminated 
	 at $J_f$, the system is freely evolved to attain a steady state. }
  \label{proto_phi_nv_dw10_ss}
\end{figure}

To study the non-equilibrium SDW-SSS quantum phase transition, we employ
the following linear quantum quench protocol
\begin{equation}
  J(t) = J_i + \frac{(J_c - J_i)}{\tau_Q}(t + \tau_Q).  
\end{equation}
With this protocol, we have, $J(-\tau_Q) = J_i$ and  $J(0) = J_c$. For our 
study we take $J_i = 0.02U$ and $J_f = 0.11U$, and fix the chemical 
potential $\mu = 0.17U$. The value of the $\mu$ is chosen so that it 
corresponds to  the tip of the SDW(1,0) lobe and $J_c = 0.067U$.
Thus, at $t = -\tau_Q$, the system is in the SDW(1,0) phase, and 
at $t = t_f$, it is in the SSS phase when the quantum quench ends.
The quench protocol employed in our simulations is shown 
in the Fig.~\ref{proto_phi_nv_dw10_ss}(b), for $\tau_Q = 100$. Once the quench 
is terminated, we let the system evolve freely so that the order parameters 
attain steady values.

%%%%%%%%%%%%%%%%%%%%%%%%%%%%%%%%%%%%%%%%%%%%%%%%%%%%%%%%%%%%%%%%%%%%%%%%%%%%%
%%%%%% Subsubsection: $\Phi$                                          %%%%%%%
%%%%%%%%%%%%%%%%%%%%%%%%%%%%%%%%%%%%%%%%%%%%%%%%%%%%%%%%%%%%%%%%%%%%%%%%%%%%%

\subsubsection{$|\Phi|$}
In the Fig.~\ref{proto_phi_nv_dw10_ss}(a), we show the temporal evolution of 
$|\Phi|$ and $N_v$ during the quantum quench for $\tau_Q = 100$. At the 
initial stages of the quantum quench, the SF order parameter is small 
$|\Phi|\approx 10^{-4}$ as the system passes through the SDW(1,0) phase 
domain. It is to be noted that, at equilibrium, $|\Phi|$ is zero in an 
insulating phase like SDW. However, in the quench dynamics it has a small 
value as the equilibrium solution is augmented with density and phase 
fluctuations during the initial state preparation. The value of $|\Phi|$ 
remains small after crossing the critical value $J_c$, at time $t=0$, till 
the time $\hat{t}$. After time $\hat{t}$ there is an initial exponential 
increase in the value of $|\Phi|$. And, this is discernible from the inset
plot in Fig.~\ref{proto_phi_nv_dw10_ss}(a). 
At a later time,
a damped oscillatory trend sets in while the average value tend towards 
the steady state value. For the present case, the quantum quench is 
terminated at $t_f = \tau_Q$, and then, we let the system evolve freely to 
attain a steady state. On the termination of the quantum quench, $|\Phi|$ 
stops increasing, and settles to its steady state value.

\begin{figure}[ht]
  \includegraphics[width = 8.5cm]{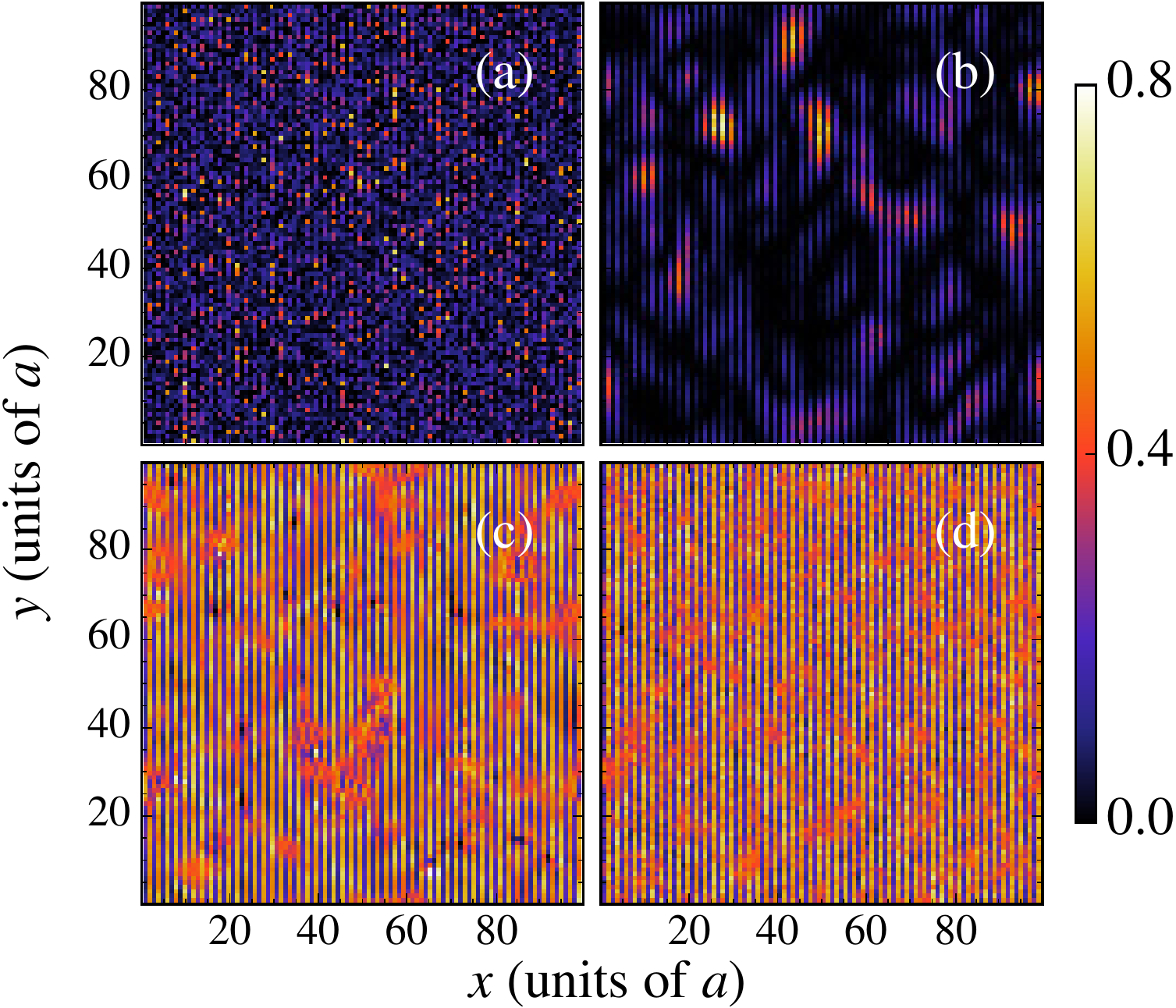}
	\caption{Snapshots of $|\phi_{p,q}|$ at certain times 
	   for $\tau_Q = 100$. 
	   Initially at $t = -\tau_Q$, $|\phi_{p,q}|$ is small as shown in (a).
	   As time progresses, the magnitude of $|\phi_{p,q}|$ increases as 
	   shown in (b). Plot (b) is at $t = \hat{t} = 28$. 
	   Domains of non-zero $|\phi_{p,q}|$ are formed at 
	   this time, and they grow in size, thereby increasing the 
	   value of $|\Phi|$. The representative progress of the phase ordering
	   and domain merging can be seen in (c) and (d). 
	   Plot (c) is at $t = 92$, while plot (d) is at $t=400$.}
  \label{phi_snap}
\end{figure}

We present snapshots of $|\phi_{p,q}|$ at important time instants during the 
quench dynamics in the Fig.~\ref{phi_snap}. At time $t = -\tau_Q$, as discussed
earlier and visible in Fig.~\ref{phi_snap}(a), $|\phi_{p,q}|$ has small 
value. Despite the small value the fluctuations in $|\phi_{p,q}|$ 
is evident from the figure. The delay in the transition to SSS phase 
till $\hat{t}$ as the system evolves through the impulse domain is reflected in
the value of $|\phi_{p,q}|$ shown in Fig.~\ref{phi_snap}(b). In the figure, the
emergence of few small SSS domains is visible as the system exits impulse 
domain and re-enters the adiabatic domain. This indicates there are local 
choices of the order parameter in the symmetry broken SSS phase. When the $J/U$
is further increased, as discussed earlier and shown in the 
Fig.~\ref{proto_phi_nv_dw10_ss}(a), $|\Phi|$ increases 
exponentially. At later times, the domains grow in size and phase ordering 
occurs. The representative progress of the phase ordering are shown in 
Fig.~\ref{phi_snap}(c) and (d).

%%%%%%%%%%%%%%%%%%%%%%%%%%%%%%%%%%%%%%%%%%%%%%%%%%%%%%%%%%%%%%%%%%%%%%%%%%%%%
%%%%%% Subsubsection: $N_v$                                           %%%%%%%
%%%%%%%%%%%%%%%%%%%%%%%%%%%%%%%%%%%%%%%%%%%%%%%%%%%%%%%%%%%%%%%%%%%%%%%%%%%%%

\subsubsection{$N_v$}
 The vortex density $N_v$, as mentioned earlier, is an indicator of the 
defect density in the system. And, for a linear quench it satisfies the 
scaling law $N_v \propto \tau_Q^{-d}$. At initial stages of the quantum
quench, the vortex density is high $N_v\approx 2500$. This occurs as 
vortices are imprinted due to the phase fluctuations introduced during 
the initial state preparation. The $U(1)$ global symmetry of the system is 
broken spontaneously as the quantum quench progresses and enters 
the SSS phase after crossing the phase boundary. The SF order parameter 
$|\Phi|$, then, acquires a finite value and $\phi_{p,q}$ must be phase 
coherent throughout the system. However, in the quenched system, domains are 
formed such that there is phase coherence within, but not between the 
domains. At later times, the domains grow in size through mergers. This is 
accompanied by annihilation of vortex-antivortex pairs, and reduces the 
vortex density $N_v$ in the system. As the quench is continued further,
phase coherence is established in the system, and $N_v$ approaches zero.
\begin{figure}[h] 
  \begin{center}
  \includegraphics[width = 8.9cm]{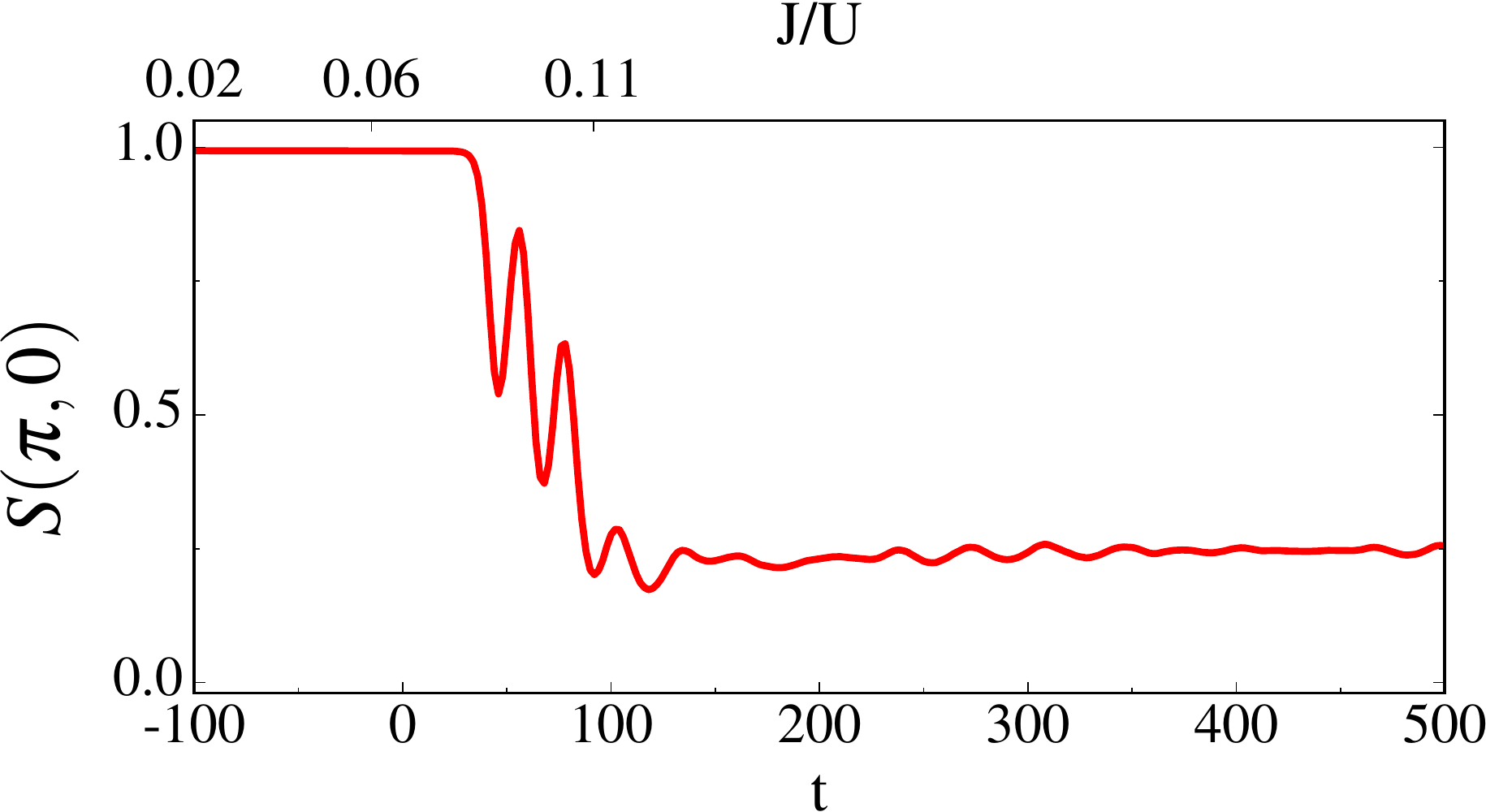}
   \caption{$S(\pi,0)$ as a function of time, for $\tau_Q = 100$. 
	  Since the SDW phase has crystalline property, $S(\pi,0)$ equals 
	  unity in this phase. In the SSS phase, the solidity persists, 
	  but there is also superfluid character in the system, and 
	  $S(\pi,0)$ deviates from unity. 
	   The hopping parameter $J/U$ corresponding to time $t$
	   is shown on the top axis. After the 
	   quench is terminated at $J_f = 0.11U$, the system is freely 
	   evolved with $J = J_f$.} 
  \label{dw_ss_sfpz}
  \end{center}
\end{figure}
Another observable which is an indicator of the SDW-SSS transition is the 
structure factor $S(\pi,0)$. The evolution of the $S((\pi,0)$ during the
quantum quench dynamics is shown in Fig.~\ref{dw_ss_sfpz}. Like 
in the temporal evolution of $|\Phi|$, there is a significant change in the 
value of $S((\pi,0)$ around the time $\hat{t}$. We observe that $S(\pi,0)$ 
deviates from unity at $\hat{t}$, when few finite sized SF domains 
develops in the system. The domains display large amplitude oscillations 
while it continues to decay exponentially. Once the quantum quench is 
terminated and the system is in the free evolution stage, $S(\pi,0)$ exhibits 
small amplitude oscillations about the steady state value.

\begin{figure}[ht]
  \includegraphics[width = 8.5cm]{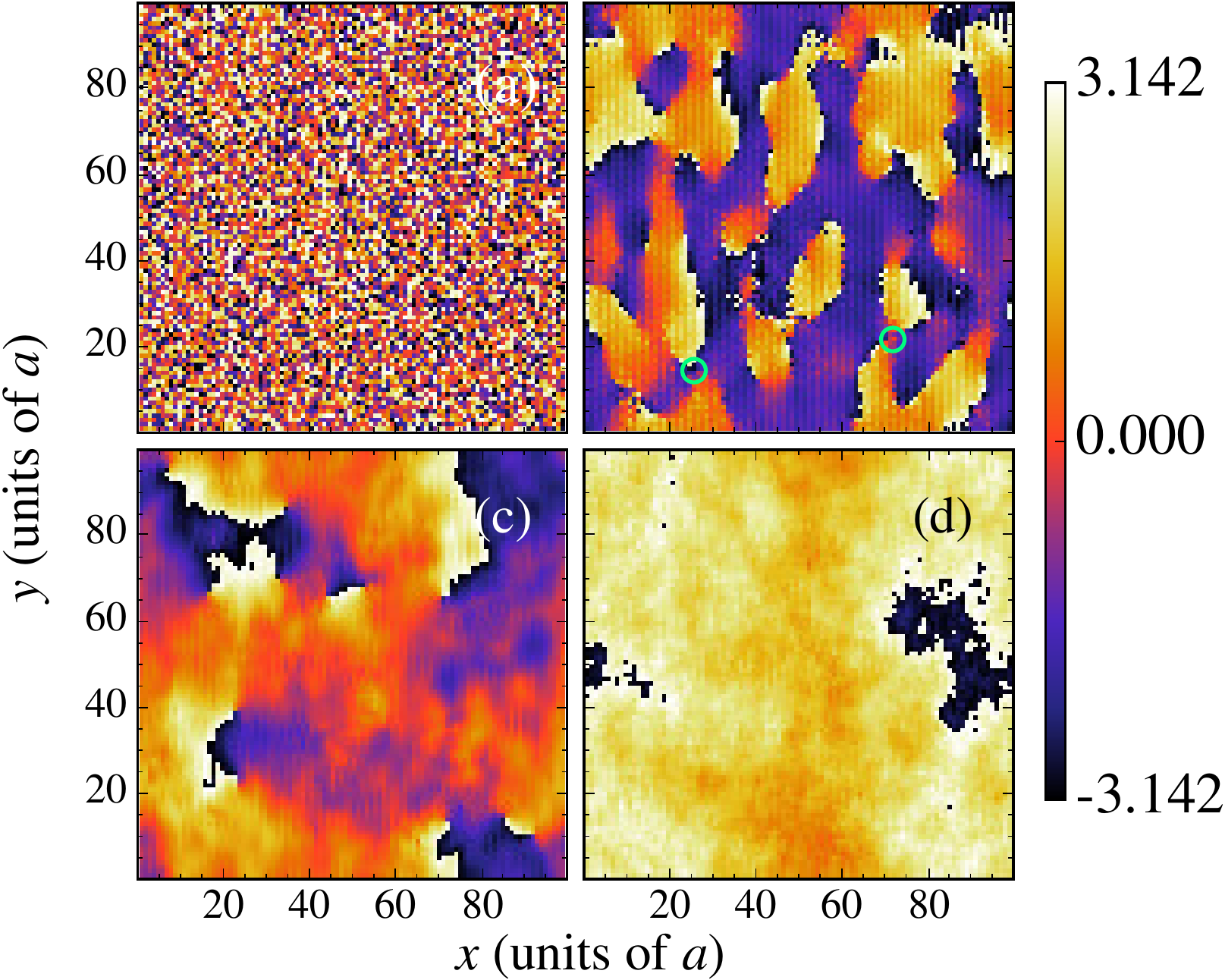}
  \caption{Snapshots of the phase of $\phi_{p,q}$ at 
	   certain times for 
	   $\tau_Q = 100$. Initially at $t = -\tau_Q$, the vortex density is 
	   high, as shown in (a), owing to the phase fluctuations imprinted to
	   the system. The domains of uniform phase are formed 
	   after the system crosses the critical point, as shown in (b), which
	   is at $t = \hat{t} = 28$. As an illustration of the
	   vortices in the system, we show two of them by the green circles.
	   There is phase ordering process which 
	   increases the size of the domains and eventually system retains 
	   an almost uniform phase, as shown in (c) and (d). 
	   Plot (c) is at $t = 92$ and (d) is at $t = 400$.}  
  \label{phphi_snap}
\end{figure}
 
 To visualize the vortices in the system, the snapshots of the phase of the 
SF order parameter $\phi_{p,q}$ are shown in Fig.~\ref{phphi_snap}. In the 
figure, the location of a vortex (antivortex) is identified as the point
around which the phase of $\phi_{p,q}$ change by $2\pi$ along anti-clockwise
(clockwise) direction. At the beginning of the quantum quench 
$t = -\tau_Q$, as seen 
from Fig.~\ref{phphi_snap}(a), many vortices are present in the system. 
The appearance of the SF domains after the system crosses the critical point 
is visible in the Fig.~\ref{phphi_snap}(b), which shows the phase of 
$\phi_{p,q}$ at $t=\hat{t}=28$. Two features are discernible in the figure.
First, the phase coherence within each of the domains are indicated by the 
near uniform color. And, second, the presence of vortices at the points 
where more than two domains meet. The reduction in the number of vortices
(antivortices) through vortex-antivortex annihilation is visible when 
we compare the plots at later times shown in Fig.~\ref{phphi_snap}(c-d).
From the sequence of the plots, post $\hat{t}$, the consolidation of the 
domains  size or phase ordering in the system is quite prominent. As seen
from Fig.~\ref{phphi_snap}(d) at later times, the system has almost uniform
phase.
\begin{figure}[ht]
  \includegraphics[width = 7.5cm]{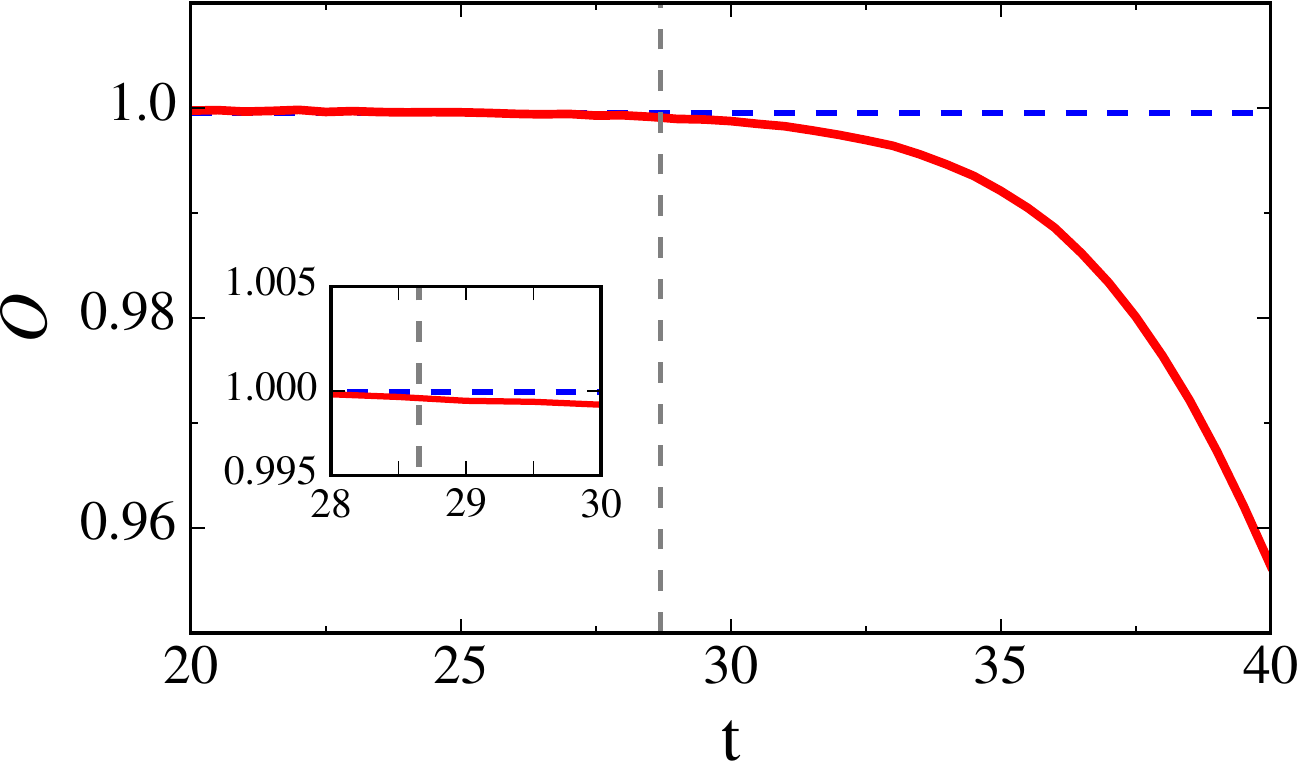}
	\caption{Plot of the overlap measure $O(t)$, defined in 
	         Eq.(\ref{olapJqnch}), to locate $\hat{t}$. The red curve 
		 indicates the overlap, while the vertical grey dashed line 
		 indicates the time $\hat{t}$
	         at which $O(t)$ begins to deviate from unity. The inset zooms 
		 into the region near $\hat{t}$. The dashed blue line is a 
		 visual guide to indicate the unity on the ordinate axis.}
  \label{overlap}
\end{figure}

%%%%%%%%%%%%%%%%%%%%%%%%%%%%%%%%%%%%%%%%%%%%%%%%%%%%%%%%%%%%%%%%%%%%%%%%%%%%%
%%%%%% Subsubsection: Protocol to locate $\hat{t}$                    %%%%%%%
%%%%%%%%%%%%%%%%%%%%%%%%%%%%%%%%%%%%%%%%%%%%%%%%%%%%%%%%%%%%%%%%%%%%%%%%%%%%%

\subsubsection{Protocol to locate $\hat{t}$}
 The scaling laws are applicable at the transition time $\hat{t}$ and hence
it is important to identify $\hat{t}$ accurately. To locate $\hat{t}$,
we compute the overlap of the wavefunction of the system at $t>0$ with
the wavefunction at $t=0$ as
\begin{align}
   O(t) = |\langle{\Psi(0)}\Ket{\Psi(t)}|.
   \label{olapJqnch}
\end{align}
As indicated earlier, the quantity $O(t)$ should be equal to unity
as long as the system is in the impulse regime or $t\leqslant \hat{t}$.
This follows as the state of the system remains frozen in the impulse
regime. This is because the state maximally picks up a phase factor, when it
is in the impulse domain\cite{damski_06}. At $\hat{t}$, when it passes from
the impulse to adiabatic regime, the overlap shall start to deviate from
unity. Thus, computing the $O(t)$, we locate the $\hat{t}$ by noting the
time at which it deviates from unity. As an example, 
Fig.~\ref{overlap} shows a generic plot of $O(t)$ around $\hat{t}$.

%%%%%%%%%%%%%%%%%%%%%%%%%%%%%%%%%%%%%%%%%%%%%%%%%%%%%%%%%%%%%%%%%%%%%%%%%%%%%
%%%%%% Subsubsection: Critical exponents and scaling relations        %%%%%%%
%%%%%%%%%%%%%%%%%%%%%%%%%%%%%%%%%%%%%%%%%%%%%%%%%%%%%%%%%%%%%%%%%%%%%%%%%%%%%

\subsubsection{Critical exponents and scaling relations}
\begin{figure}[ht]
  \includegraphics[height = 8cm]{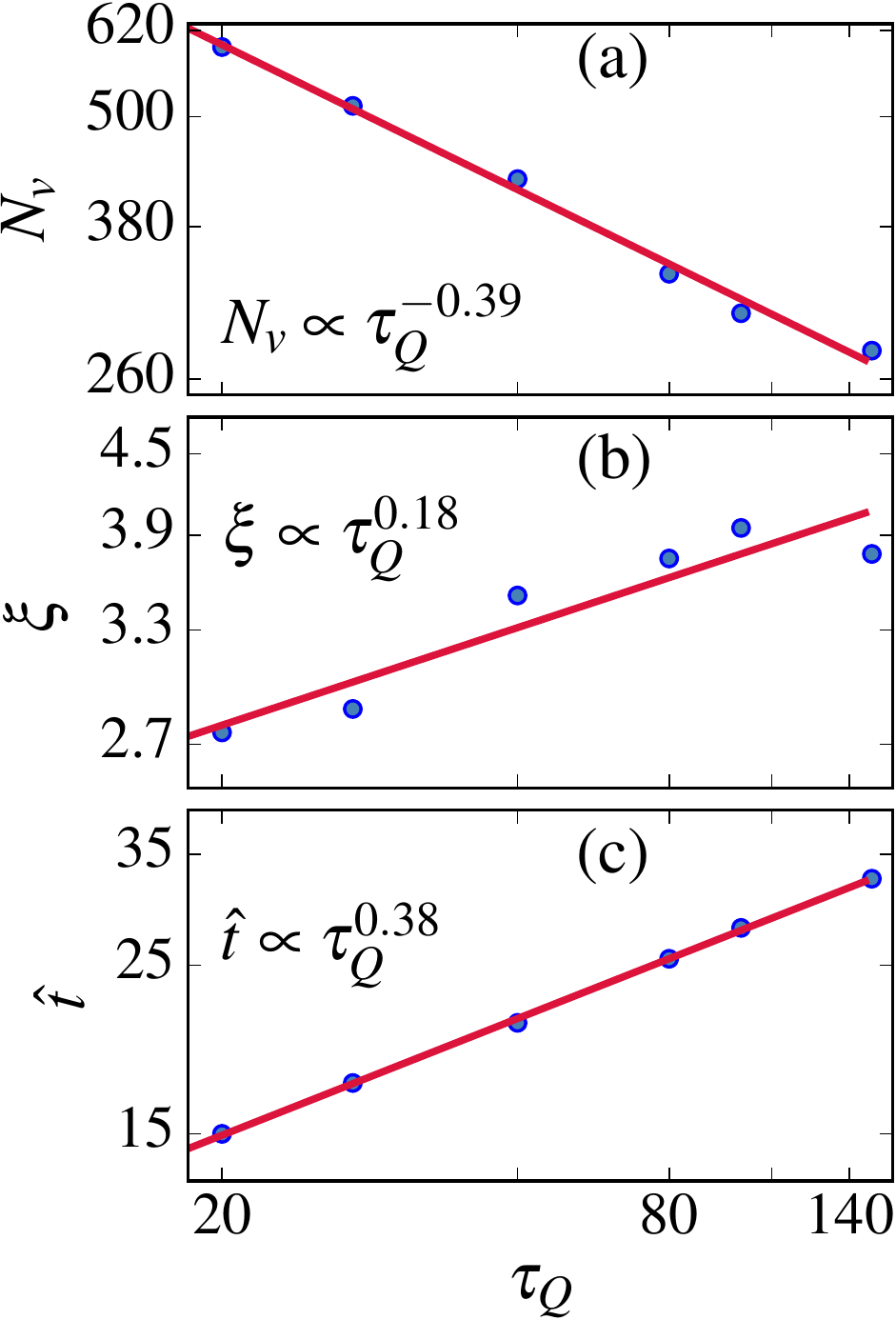}
  \caption{Power law scaling of $N_v$, $\xi$ and 
	    $\hat{t}$ with respect to $\tau_Q$. The exponent $d = 0.39$, 
	    $b = 0.18$ and the $\hat{t} \propto \tau_Q^{0.38}$. The obtained 
	    exponents $b$ and $d$ approximately satisfy the scaling relation 
	    $d = 2b$.}
  \label{nv_scal}
\end{figure}

 To determine the critical exponents $b$ and $d$, we do a series of 
computations over a range of $\tau_Q$. The values of $N_v$ and $\xi$ 
obtained as a function of $\tau_Q$ are shown in Fig.~\ref{nv_scal}(a-b). 
The critical exponents $b$ and $d$ obtained from these are $0.18$ and $0.39$, 
respectively. And, they approximately satisfy the scaling relation $d=2b$ 
obtained from Eq. (\ref{crit_exp}). 
The KZ mechanism also predicts the scaling law for the transition time
\begin{equation}
  \hat{t} \propto \tau_Q ^{\frac{\nu z}{1+\nu z}},
\end{equation}
and a plot of the $\hat{t}$ as a function of $\tau_Q$ is shown in 
Fig.~\ref{nv_scal}(c). Comparing with the expression of $b$ in 
Eq. (\ref{crit_exp}), we get $z \approx 2$.
We have also studied the dynamics of the SDW(2,0)-SSS 
quantum phase transition. The quench protocol is similar to the case of
SDW(1,0)-SSS transition. For a given $\tau_Q$, we observe
larger $\hat{t}$ for the SDW(2,0)-SSS transition than the 
SDW(1,0)-SSS transition. From the results we get critical
exponents as $b = 0.19$, and $d = 0.39$. Thus the critical exponents obey 
the scaling relation $d = 2b$.

%%%%%%%%%%%%%%%%%%%%%%%%%%%%%%%%%%%%%%%%%%%%%%%%%%%%%%%%%%%%%%%%%%%%%%%%%%%%%
%%%%%% Subsection: Tilt angle quenching                               %%%%%%%
%%%%%%%%%%%%%%%%%%%%%%%%%%%%%%%%%%%%%%%%%%%%%%%%%%%%%%%%%%%%%%%%%%%%%%%%%%%%%

\subsection{Quenching $\theta$}
 The dipolar gases in optical lattice offer an unique possibility
of tuning the tilt angle $\theta$, thereby control the anisotropy of the 
dipolar interactions. This leads to a rich equilibrium phase diagram of the 
system \cite{zhang_15, bandyopadhyay_19}. The quenching of the 
tilt angle $\theta$ can be done in experiments by changing the orientation of 
the external magnetic field \cite{baier_16}. And, when $\theta$ is quenched 
across $\theta_M$ in our system, the dipolar interaction along the $y$-axis 
changes from repulsive to attractive. This leads to a first order quantum phase 
transition from the striped to checkerboard order. We study this first order 
phase transition from the perspective of the KZ mechanism. And, the same has 
been attempted in a few previous works. But, not all of them show the power-law 
scaling of the defect density with the quench rate.  
In \cite{shimizu_dwsf_18}, a power law dependence is obtained
for the topological defects across the density wave-to-superfluid transition,
while in \cite{coulamy_17}, an exponential saturation of the defects is 
observed for the dynamics of a quantum search. In \cite{qiu_20}, the authors 
have considered quantum quench across the polar and the anti-ferromagnetic 
phases in spinor condensates, and have shown the presence of the adiabatic and
the impulse domains in the evolution. They report the power-law scaling of the 
tuning parameter in the quench. It is thus pertinent to check for the presence 
of the impulse and adiabatic regimes in the dynamics of striped to checkerboard 
transition, and the associated power-law scalings.

%%%%%%%%%%%%%%%%%%%%%%%%%%%%%%%%%%%%%%%%%%%%%%%%%%%%%%%%%%%%%%%%%%%%%%%%%%%%%
%%%%%% Subsection: Equilibrium aspects                                %%%%%%%
%%%%%%%%%%%%%%%%%%%%%%%%%%%%%%%%%%%%%%%%%%%%%%%%%%%%%%%%%%%%%%%%%%%%%%%%%%%%%

\subsubsection{Equilibrium ground state}
\label{sss_cbss_eqlb}
\begin{figure}[ht]
  \includegraphics[width = 8.5cm]{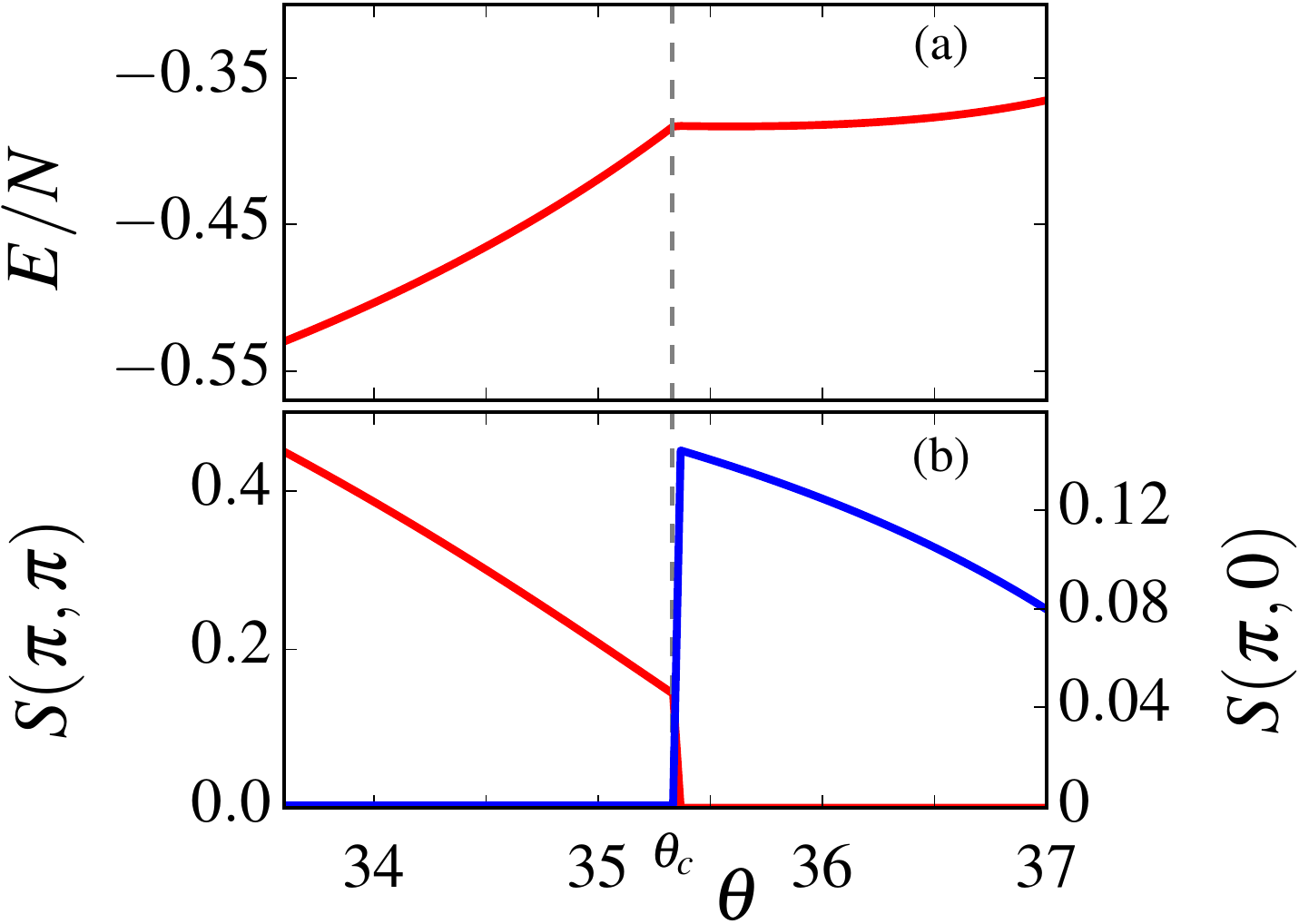}
     \caption{(a)Energy per particle of the ground state of 
	      the dipolar system
              as a function of the tilt angle $\theta$. There is a 
	      discontinuity in the first derivative of the energy around 
	      $\theta_c = 35.3\degree$. The $\theta$ values are denoted 
	      in units of degrees. (b) Structure factors $S(\pi,\pi)$ (red) and 
	      $S(\pi,0)$ (blue) as a function of the tilt angle $\theta$. Both 
	      the quantities show a discontinuous jump at $\theta_c$.}
     \label{en_sfpp}
\end{figure}

  For the quench dynamics, we fix the parameters of the system as $J=0.05U$, 
and $\mu=0.75U$. The corresponding critical tilt angle $\theta_c$ at which the 
SSS-CBSS quantum phase transition occurs is $35.3\degree$, and
thus $\theta_c \approx \theta_M$. As a first step to examine the quench 
dynamics we characterise the equilibrium ground state of the system as a 
function of $\theta$. In particular, the kink in the energy per particle $E/N$ 
of the ground state at the $\theta_c$, as shown in Fig.~\ref{en_sfpp}(a), 
indicates a first order quantum phase transition. The system has striped order 
along $y$-axis (SSS phase) and checkerboard order for $\theta > \theta_c$ and  
$\theta < \theta_c$, respectively. The two quantum phases are discernible from 
the structure factors $S(\pi,\pi)$ and $S(\pi,0)$. As shown in 
Fig.~\ref{en_sfpp}(b), $S(\pi,\pi)$ and $S(\pi,0)$ are non-zero for the phases
with checkerboard and striped order, respectively. In the neighbourhood of 
$\theta_M$, the converged solutions are sensitive to the initial state. The 
most common solution is a metastable emulsion phase, which is a superposition 
of the checkerboard and striped orders \cite{zhang_15, bandyopadhyay_19}. So, 
to identify the equilibrium ground state, we use different initial states and 
consider the converged solution with the lowest energy as the ground state.

%%%%%%%%%%%%%%%%%%%%%%%%%%%%%%%%%%%%%%%%%%%%%%%%%%%%%%%%%%%%%%%%%%%%%%%%%%%%%
%%%%%% Subsection: Equilibrium aspects                                %%%%%%%
%%%%%%%%%%%%%%%%%%%%%%%%%%%%%%%%%%%%%%%%%%%%%%%%%%%%%%%%%%%%%%%%%%%%%%%%%%%%%

\subsubsection{SSS-CBSS quench dynamics}
The quench protocol to study the dynamics of the SSS-CBSS  quantum phase
transition is 
\begin{equation}
  \theta(t) = \theta_i - \frac{(\theta_i - \theta_c)}{\tau_Q}(t + \tau_Q).  
  \label{thta_qnch_protcol}
\end{equation}
As mentioned earlier, we consider $J = 0.05U$, and $\mu = 0.75U$ for the 
quench dynamics. Using the above linear quenching protocol, 
we have $\theta(-\tau_Q) = \theta_i$, and $\theta(0) = \theta_c$. The 
simulations are performed with $\theta_i = 37\degree$ and 
$\theta_f = 32\degree$, 
and consider a wide range of the quench times $\tau_Q$. The choice of the 
parameters are based on the ground state phase diagram of the model 
Hamiltonian of the system \cite{bandyopadhyay_19}. Like in the $J$ quenching, 
the quantum fluctuations essential to initiate quantum phase transition are 
incorporated through the addition of noise to the coefficients in the 
Gutzwiller wave function. 

\begin{figure}[ht]
  \includegraphics[width = 8.5cm]{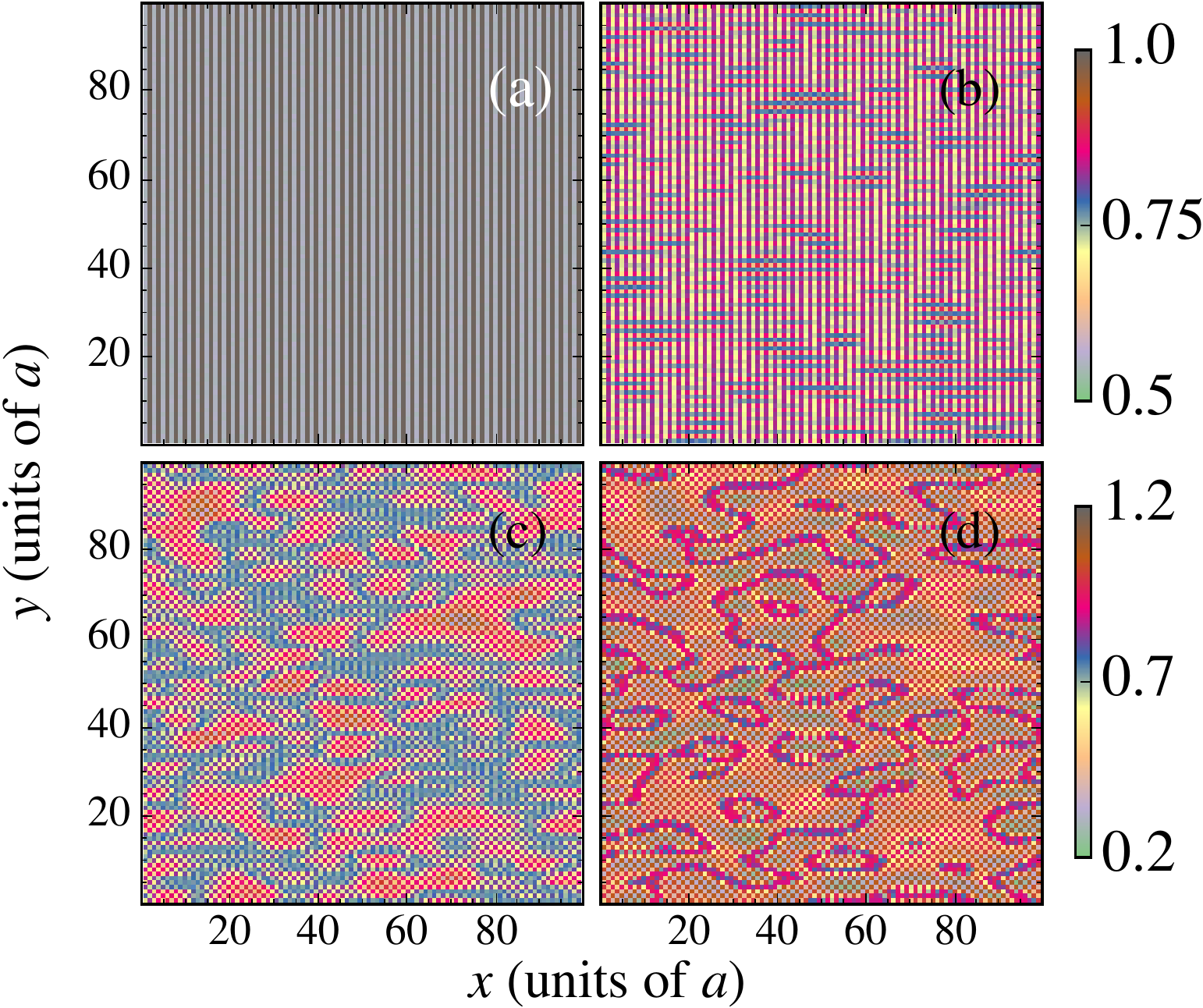}
    \caption{Snapshots of the density at certain times in 
	    the time evolution across SSS-CBSS transition, for $\tau_Q = 20$.  
 	    The subplot (a) corresponds to the initial state at time 
	    $t = -\tau_Q$, (b) corresponds to the state at time 
	    $\hat{t} = 44$, (c) denotes the density profile at $t = 50$, and 
       	    (d) corresponds to the density profile at the end of the quench.
	    We observe domains of checkerboard order emerging at $\hat{t}$.
	    The density ordering can be noticed from subplots (b)-(d) in the 
	    figure.} 
    \label{den_sss_cbss}
\end{figure}

 The density profiles of the system from one of the simulations at different
stages of the evolution are shown in Fig.~\ref{den_sss_cbss}. Initially, at 
the beginning of the quench ( $t = -\tau_Q$), the density has striped order.
This is shown in the  Fig.~\ref{den_sss_cbss}(a). The striped order is, then,
altered after the system crosses the critical tilt angle $\theta_c$. The 
resulting density deformations emerge as  patches of checkerboard domains and
the system appears like shown in Fig.~\ref{den_sss_cbss}(b). These domains 
grows in size and the density ordering progresses. At late times, there are 
large domains of checkerboard order separated by a ``network'' of remnant
striped order as seen in Fig.~\ref{den_sss_cbss}(d).

%%%%%%%%%%%%%%%%%%%%%%%%%%%%%%%%%%%%%%%%%%%%%%%%%%%%%%%%%%%%%%%%%%%%%%%%%%%%%
%%%%%% Subsection: Equilibrium aspects                                %%%%%%%
%%%%%%%%%%%%%%%%%%%%%%%%%%%%%%%%%%%%%%%%%%%%%%%%%%%%%%%%%%%%%%%%%%%%%%%%%%%%%

\subsubsection{Scaling of $\hat{t}$}
\label{scal_that_ss_cb}
\begin{figure}[ht]
  \includegraphics[height = 5.5cm]{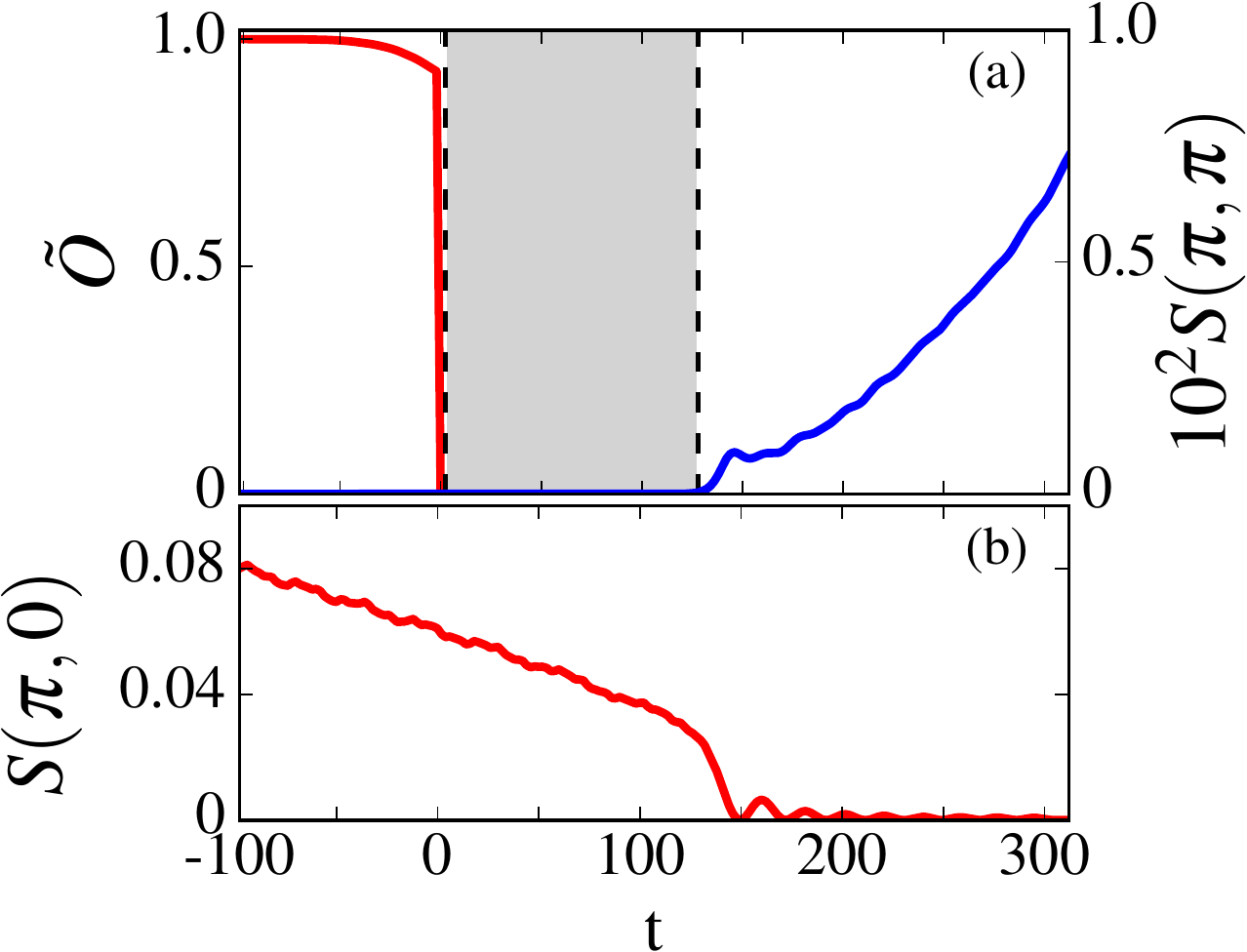}
  \caption{Overlap $\tilde{O}$ between the ground state and the
	   dynamically evolved state, and the structure factor $S(\pi,\pi)$ as 
	   a function of time is shown in (a). The grey shaded region denotes 
	   the impulse domain in the dynamics. 
	   The time instant $\hat{t}$, which marks the end of the impulse 
	   domain, is when the $S(\pi,\pi)$ becomes non-zero. In (b), we show
	   the $S(\pi,0)$ as a function of time. There is a decreasing trend
	   in the $S(\pi,0)$, and near $\hat{t}$, the decay is steeper.}
  \label{olap_sfpp}
\end{figure}

 To identify the end of the adiabatic domain and onset of the impulse domain 
of the phase transition, we use the overlap $\tilde{O}$. 
It is the overlap of the quenched state with the equilibrium ground state 
having same $\theta$ value. The trend of $\tilde{O}(t)$ is as shown in the 
Fig.~\ref{olap_sfpp}(a) with the red colored plot. At the early stages, the 
overlap is close to unit value and this is representative of the adiabatic 
evolution. As the quench progresses, it deviates from unity, and 
becomes zero. This marks the beginning of the impulse domain. Then, the ground 
state of the system, which has a checkerboard ordering, does 
not show good fidelity with the quenched state. The time instant when the 
overlap becomes zero depends on the quench rate, but it is located near 
$\theta_c$. We expect $\tilde{O}(t)$ to be non-zero when the system leaves 
impulse domain and re-enters adiabatic domain. However, post impulse domain, 
the state is populated with topological defects. And, thus this state does not 
show good fidelity with the corresponding equilibrium ground state. So, 
$\tilde{O}(t)$ is not a good indicator of the post-impulse adiabatic domain.

 To mark the end of the impulse domain, we use $S(\pi,\pi)$ as an indicator.
Its trend with time is as shown by the blue color plot in the  
Fig.~\ref{olap_sfpp}(a). The $S(\pi,\pi)$ is zero in the SSS phase, but
finite in the CBSS phase. Thus, the emergence of the checkerboard order is
when $S(\pi,\pi)$ becomes non-zero. In analogy with the $J$ quenching, we 
mark this time instant as $\hat{t}$. In other words, there is a time delay 
in the emergence of the checkerboard order after crossing $\theta_c$. This 
delay implies the extent of the frozen or the impulse domain of the quench
dynamics. In Fig.~\ref{olap_sfpp}(a), the grey region denotes the impulse 
domain. As the quench dynamics progresses, the $S(\pi,0)$ 
decreases and this trend is discernible from the plot in the
Fig.~\ref{olap_sfpp}(b). This trend continues after crossing the $\theta_c$ at 
$t=0$, and it arises from the density fluctuations present in the impulse 
domain. Similar observations were reported in the previous works
\cite{biroli_10, jeong_19} describing a coarsening dynamics in 
the impulse domain. Thus, even though the system displays a striped pattern in 
the impulse domain, the fluctuations cause local density changes, thereby 
changing the $S(\pi,0)$. However, after crossing $\hat{t}$, when the system 
enters the adiabatic domain, the $S(\pi,0)$ decreases faster. It becomes zero 
after some characteristic oscillations as the checkerboard order sets in as the 
ground state. Another point to note is that there is a finite time interval 
after $\hat{t}$, where both the $S(\pi,0)$ and $S(\pi,\pi)$ are 
non-zero. That is, post $\hat{t}$, there are domains of checkerboard order with
the striped order as the background, it resembles the emulsion phase 
\cite{bandyopadhyay_19}. However, as the quench progresses further, the striped
pattern diminishes, and the $S(\pi,0)$ becomes zero.

\begin{figure}[ht]
  \includegraphics[height = 7.5cm]{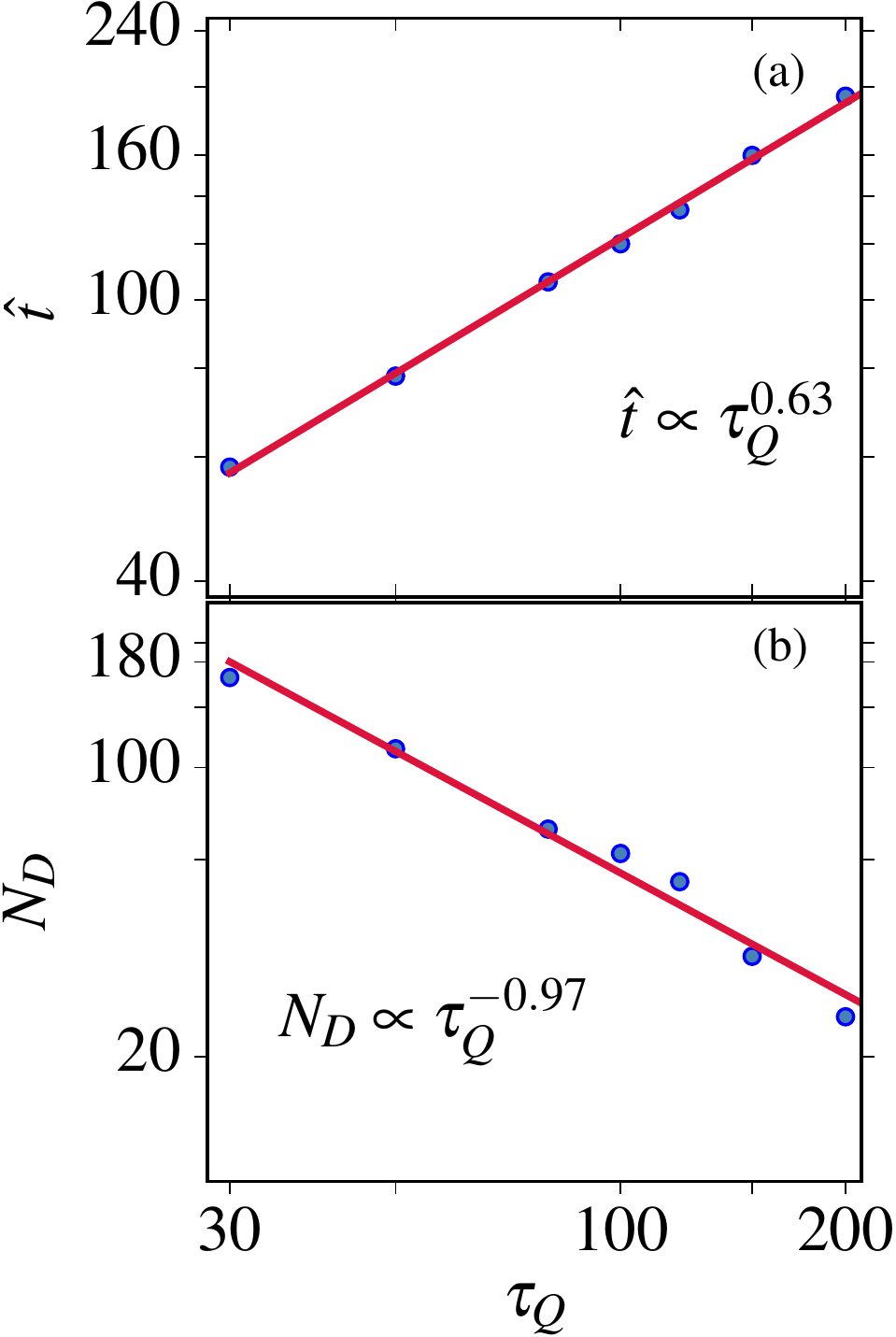}
	\caption{Scaling of $\hat{t}$ and  $N_D$  with $\tau_Q$, for the 
	SSS-to-CBSS phase transition. The power-law scaling of $\hat{t}$
	is discernible in (a), and the exponent is $0.63$. 
	In (b), the scaling of $N_D$ is presented and the exponent is 
	$d = 0.97$.}
  \label{sss_cbss_scal}
\end{figure}

The scaling of the $\hat{t}$ with $\tau_Q$ is shown 
in the Fig.~\ref{sss_cbss_scal}(a). The plot shows the power-law nature
of $\hat{t}$ with $\tau_Q$, and the corresponding critical exponent is 0.63. 
That is, for the SSS-CBSS quantum phase transition
$\hat{t} \propto \tau_{Q}^{0.63}$.

%%%%%%%%%%%%%%%%%%%%%%%%%%%%%%%%%%%%%%%%%%%%%%%%%%%%%%%%%%%%%%%%%%%%%%%%%%%%%
%%%%%% Subsection: Equilibrium aspects                                %%%%%%%
%%%%%%%%%%%%%%%%%%%%%%%%%%%%%%%%%%%%%%%%%%%%%%%%%%%%%%%%%%%%%%%%%%%%%%%%%%%%%

\subsubsection{Scaling of the Number of domains}
\label{scal_ndom_ss_cb}
  
\begin{figure}[ht]
  \includegraphics[width = 8.5cm]{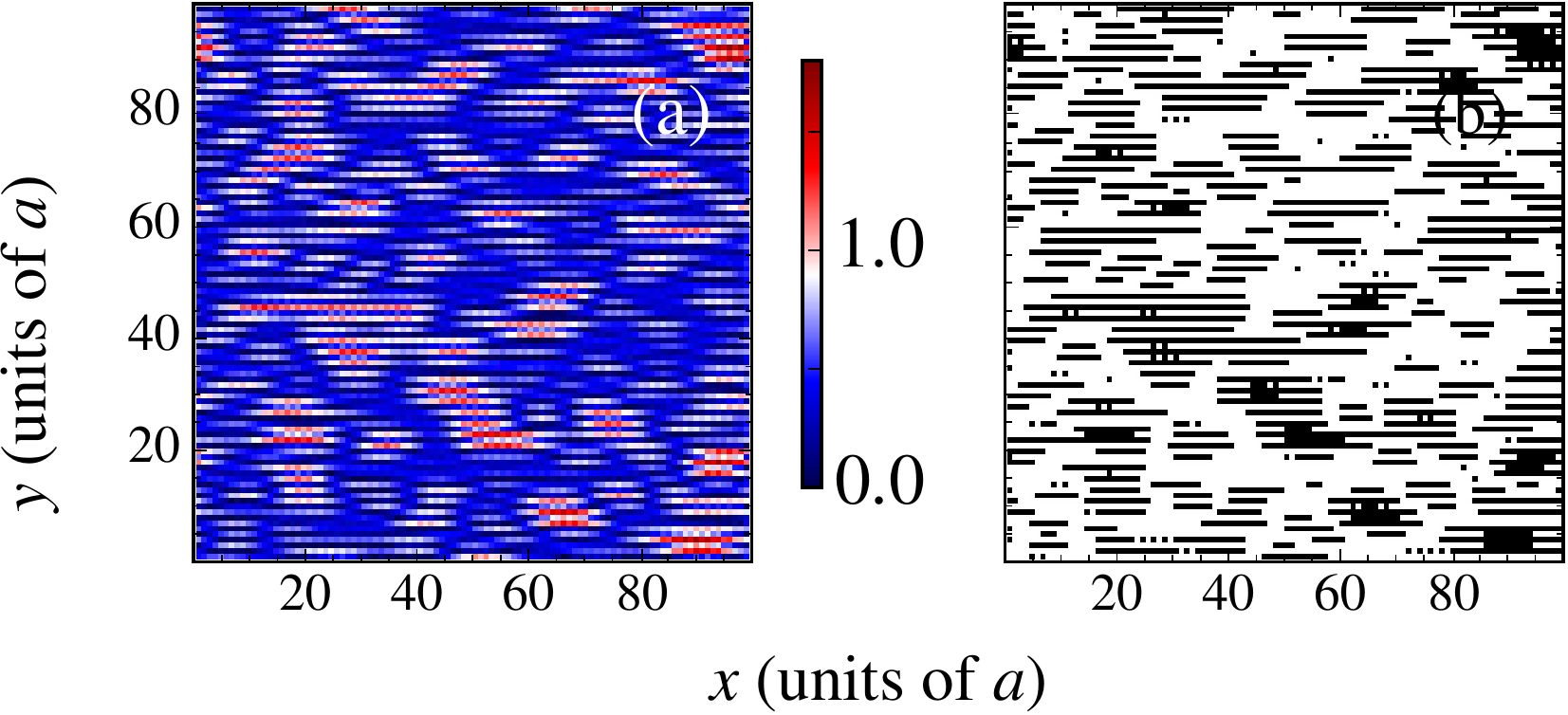}
     \caption{(a) Plot of the density contrast order parameter 
	$N_{i,j}$. The regions with the values of this order parameter smaller 
	than a certain
	threshold value correspond to the 
	striped phase, while those with larger values denote the domains of 
	checkerboard order that are formed in the time dynamics.
	(b) The plot of the distribution of the binary values $0$ and $-1$, 
	obtained after applying the threshold to the plot (a). 
	The black colored domains indicate those
	of label 0, while the white regions indicate the domains of label $-1$.
	The domains of the label 0 are counted numerically.}
     \label{nij_cbss_sss}
\end{figure}

 One key result of KZM is the scaling of the number of topological defects
like vortices $N_v$ or domains $N_D$. The scaling relation of $N_v$ is given
in Eq. (\ref{nv-d}). However, in the SSS-CBSS transition $N_v$ is not a good 
choice as the transition is between phases with density or diagonal orders. 
So, we consider the  number of domains $N_D$, and study the scaling relation 
with the $\tau_Q$. As a first step, to differentiate the SSS and CBSS domains
we introduce a density contrast order parameter at the $(i,j)$ lattice site
as
\begin{eqnarray}
   N_{i,j} = \Big|4\langle n_{i,j} \rangle - \big(
             \langle n_{i+1,j} \rangle + \langle n_{i-1,j} \rangle + 
             \langle n_{i,j+1} \rangle   
            + \langle n_{i,j-1} \rangle \big)\Big|. \nonumber 
   \label{nij_eqn}
\end{eqnarray}
The value of this order parameter is ideally $4 (n_A - n_B)$
and $2 (n_A - n_B)$ for the  checkerboard and striped order, respectively.
Hence it serves to contrast between the two ordered phases.
As the phases of the quenched states are out-of-equilibrium,  $N_{i,j}$ may
have a range of values. And, a representative plot is shown in 
Fig.~\ref{nij_cbss_sss}(a).
To identify the domains we define a threshold value $\epsilon$ of the
$N_{i,j}$ and it is set equal to the $N_{i,j}$ of the striped domain.
Then the regions with $N_{i,j} > \epsilon$ are considered
as CBSS phase and those with lower value as SSS. The lattice sites are, then,
labeled as 0 and $-1$ for the CBSS and SSS phases, respectively. To count
the number of domains we use percolation analysis, and the computational 
algorithms are described in previous works \cite{hoshen_76, shitara_17}. In
the present work, however, we have developed an algorithm based on 
linked-lists \cite{sable_21}. With this approach a single scan across the 
system is sufficient to count the domains and label them uniquely. To validate 
the algorithm, we calculate the $N_D$ for the SDW(1,0)-SSS phase transition. 
And, we obtain the critical exponent $d = 0.40$, which is in good agreement 
with the value $d = 0.39$ obtained from the scaling of $N_v$ as shown
earlier in the Fig.~\ref{nv_scal}. 

 In the case of the SSS-CBSS transition, for each value of $\tau_Q$ we do an 
averaging for better statistical description. To do this we select fifteen 
members at random from the ensemble and then, partition these into three sets. 
From each set we obtain a value of $\epsilon$ by taking average of the 
prominent striped domains. Using each of the three values we calculate the 
ensemble average of $N_D$ and take their mean. This statistical averaging
provides good representation of the variation of $\epsilon$ between the 
members of the ensemble. Then, the scaling of $N_D$ at 
$\hat{t}$ with the quench rate $1/\tau_Q$ is analysed and the results are 
shown in Fig.~\ref{sss_cbss_scal}(b). We obtain a power-law scaling with the 
quench rate, as expected from the KZM. The critical exponent is 0.97. Thus, the quantum phase transition from the 
striped supersolid to checkerboard supersolid phase obeys the KZ scalings.

%%%%%%%%%%%%%%%%%%%%%%%%%%%%%%%%%%%%%%%%%%%%%%%%%%%%%%%%%%%%%%%%%%%%%%%%%%%%%%%
%%%%                           Conclusions                               %%%%% 
%%%%%%%%%%%%%%%%%%%%%%%%%%%%%%%%%%%%%%%%%%%%%%%%%%%%%%%%%%%%%%%%%%%%%%%%%%%%%%%

\section{Conclusions}\label{conclude}
 In conclusion, we have examined the quantum quench dynamics of the dipolar
Bose-Hubbard model in 2D optical lattice. The quenching of the 
hopping amplitude across the SDW(1,0) - SSS quantum phase transition 
is studied from the perspective of the KZ mechanism. Specifically,
we obtain the power-law scaling of the vortex density, and the correlation 
length of the system, at time instant $\hat{t}$, and the critical exponents
$b$ and $d$ obey the scaling relation $d = 2b$. Then, the structural quantum 
phase transition from the striped to the checkerboard supersolid phase is 
studied by quenching the tilt angle $\theta$. We show the existence of the 
adiabatic and the impulse domain in this first order phase transition.
Then, the domains of checkerboard that
gets formed are analysed, and the number of such domains present at time 
$\hat{t}$ is determined, using the domain counting algorithm based on the 
linked-lists. The number of domains is shown to obey a power-law scaling 
behaviour with the quench rate, as expected in KZ mechanism.

%%%%%%%%%%%%%%%%%%%%%%%%%%%%%%%%%%%%%%%%%%%%%%%%%%%%%%%%%%%%%%%%%%%%%%%%%%%%%%%
%%%%                           Acknowledgement                           %%%%% 
%%%%%%%%%%%%%%%%%%%%%%%%%%%%%%%%%%%%%%%%%%%%%%%%%%%%%%%%%%%%%%%%%%%%%%%%%%%%%%%

\begin{acknowledgments}
The results presented in the paper are based on the computations
using Vikram-100, the 100TFLOP HPC Cluster at Physical Research Laboratory, 
Ahmedabad, India. S.B acknowledges the support by Quantum Science and 
Technology in Trento (Q@TN), Provincia Autonoma di Trento, and the ERC Starting
Grant StrEnQTh (Project-ID  804305). R.N. further acknowledges DST-SERB for 
Swarnajayanti fellowship File No. SB/SJF/2020-21/19.
\end{acknowledgments}

%%%%%%%%%%%%%%%%%%%%%%%%%%%%%%%%%%%%%%%%%%%%%%%%%%%%%%%%%%%%%%%%%%%%%%%%%%%%%
%%%%%%%%%     Appendix                                                %%%%%%%
%%%%%%%%%%%%%%%%%%%%%%%%%%%%%%%%%%%%%%%%%%%%%%%%%%%%%%%%%%%%%%%%%%%%%%%%%%%%%

\appendix
\section{Appendix}
\label{bdg_appendix}
 Here we comment on the choice of the cutoff $\Delta$, used in generating the 
random numbers. The value of $\Delta$ is fixed by calculating the collective 
excitations of the equilibrium ground state at initial time. 
These Bogoliubov-de Gennes (BdG) modes are obtained by adding a fluctuation 
to the equilibrium ground state \cite{krutitsky_10, krutitsky_11, saito_12, 
bai_20, malakar_20}. 
\begin{equation}
	c_n^{(p,q)}(t) = \left(\bar{c}_n^{(p,q)} + \delta c_n^{(p,q)}(t)\right)
			 e^{-i\omega_0^{(p,q)} t},
\end{equation}
where $\bar{c}_n^{(p,q)}$ corresponds to the equilibrium ground state 
coefficient, and $\delta c_n^{(p,q)}(t)$ is the fluctuation added to the 
ground state. The $\omega_0^{(p,q)}$ corresponds to the energy of the 
unperturbed state at the lattice site $(p,q)$.
To obtain the collective excitations, we use the Bogoliubov approximation and  
define
\begin{equation}
	\delta c_n^{(p,q)}(t) = u_n^{(p,q)}e^{-i\omega t} + v_n^{* (p,q)}
	                        e^{i\omega t},
\end{equation}
where $\omega$ is the energy of the collective mode, and 
$(u_n, v_n)$ is the amplitude of the collective mode. The $c_n^{(p,q)}(t)$ are
then used in the dynamical Gutzwiller equations, and terms that are linear
in the fluctuations part are retained. The resulting set of equations is 
referred to as the BdG equations. After solving the equations, we obtain the 
$\delta c_n$. Once we obtain $\delta c_n$, we use it to calculate the 
$\delta \langle \hat{n} \rangle$  which correspond to the change in the 
expectation value of the density. It is given as 
\begin{equation}
  \delta \langle \hat{n}_{p,q} \rangle = \sum_{k} k \left( \bar{c}_k^{*(p,q)} 
	                           \delta c_k^{(p,q)} + \bar{c}_k^{(p,q)}
				   \delta c_k^{* (p,q)} \right).
\end{equation}				  
The order of the magnitude of the $\delta \langle \hat{n}_{p,q} \rangle$ is 
then used to fix the $\Delta$. For the transition from the SSS to CBSS phase, 
the calculation of the BdG modes yield $\delta \langle \hat{n} \rangle
\approx 10^{-2}$. Thus we accordingly fix the $\Delta \approx 10^{-4}$ so that
the $\delta \langle \hat{n} \rangle \approx 10^{-2}$.
The same approach is used in fixing the strength of the 
randomness in the density fluctuations, for the SDW-to-SSS quantum phase 
transition. Thus we fix the amount of the density fluctuations by studying the 
collective excitations of the initial equilibrium state.

\bibliography{dip_tdg}{}

\end{document}